\documentclass[a4paper,fleqn,usenatbib]{mnras}

\usepackage{aas_macros}
\usepackage{upgreek}

\usepackage{mathptmx} %This package provides Times fonts also to math exressions.

\usepackage[T1]{fontenc}
\usepackage{ae,aecompl}

\usepackage{graphicx}	% Including figure files
\usepackage{amsmath}	% Advanced maths commands
\usepackage{amssymb}	% Extra maths symbols

\usepackage[usenames,dvipsnames]{color} %Colors

\newcommand{\dd}{\mathrm{d}}

\title{Mass transfer in white dwarf-neutron star binaries}
\author[A.Bobrick, M.B.Davies \& R.P.Church]
  {Alexey~Bobrick$^1$\thanks{E-mail: alexey@astro.lu.se},
  Melvyn~B.~Davies$^1$\thanks{E-mail: mbd@astro.lu.se},
  Ross~P.~Church$^1$\thanks{E-mail: ross@astro.lu.se}\\
  $^1$ Lund Observatory, Department of Astronomy and Theoretical Physics, Box 43, SE 221-00, Lund, Sweden}

\date{Accepted XXX. Received YYY; in original form ZZZ}

\pubyear{2017}

\begin{document}
\label{firstpage}
\pagerange{\pageref{firstpage}--\pageref{lastpage}} 
\maketitle

\begin{abstract}
We perform hydrodynamic simulations of mass transfer in binaries that contain a white dwarf and a neutron star (WD--NS binaries), and measure the specific angular momentum of material lost from the binary in disc winds.  By incorporating our results within a long-term evolution model we measure the long-term stability of mass transfer in these binaries.  We find that only binaries containing helium white dwarfs with masses less than a critical mass of $M_{\rm WD,crit}=0.2\,M_\odot$ undergo stable mass transfer and evolve into ultra-compact {\it X}-ray binaries. Systems with higher-mass white dwarfs experience unstable mass transfer, which leads to tidal disruption of the white dwarf. Our low critical mass compared to the standard jet-only model of mass loss arises from the efficient removal of angular momentum in the mechanical disc winds which develop at highly super-Eddington mass-transfer rates. We find that the eccentricities expected for WD--NS binaries when they come into contact do not affect the loss of angular momentum, and can only affect the long-term evolution if they change on shorter timescales than the mass-transfer rate. Our results are broadly consistent with the observed numbers of both ultra-compact {\it X}-ray binaries and radio pulsars with white dwarf companions. The observed calcium-rich gap transients are consistent with the merger rate of unstable systems with higher-mass white dwarfs.

\end{abstract}


\section{Introduction}

\label{sec:Intro}

Binary star systems containing a white dwarf and a neutron star ({\it WD--NS
binaries} hereafter) are an important class of compact binaries. They are
thought to be among the main progenitors for ultra-compact {\it X}-ray binaries
\citep[UCXBs;][]{vanHaaften2013}, which  are low-mass {\it X}-ray binaries with orbital periods of less than an hour. WD--NS binaries are a likely source for calcium-rich gap transients \citep{Perets2010,Kasliwal2012}, a recently discovered class of optical supernova-like events characterised by luminosities in the gap between those of novae and supernovae. A subset of WD--NS binaries may power a special type of long duration gamma-ray burst characterised by the absence of an accompanying supernova \citep{King2007}. These systems are good sources for space-based gravitational wave detection missions, e.g. eLISA \citep{2014arXiv1407.3404A}. They likely form the second-largest fraction of doubly compact binaries \citep{Nelemans2001}. Due to their high mass-loss rates and the nucleosynthesis happening in the disc  they may potentially have implications for galactic chemistry \citep{Margalit2016}. These systems have a close connection to  millisecond pulsars, which are rapidly spinning magnetic neutron stars \citep{2010HiA....15..121W}, and have a broader relevance to binary stellar evolution for having undergone a common envelope phase \citep{2010NewAR..54...87N}.

WD--NS binaries come into contact due to gravitational wave emission. Depending on the masses of the two components, this leads to stable or unstable mass transfer. The critical donor mass, which separates the stable systems with low-mass white dwarfs from unstable systems with higher-mass white dwarfs, depends sensitively on how much mass and angular momentum is lost during mass transfer \citep{Webbink1985}.

Stable mass transfer is known to take place for the white dwarfs of sufficiently low mass, which are observed as ultra-compact {\it X}-ray binaries. 
At present there are $14$ such sources confirmed \citep{2012A&A...543A.121V,Sanna2016}, all of them containing ultra-light donors of masses below $0.07\,M_\odot$. Their evolution is characterised by a gradual decrease in the mass-transfer rate and luminosity \citep{2012A&A...537A.104V}.

Unstable mass transfer takes place for white dwarfs of sufficiently high mass. Accelerated mass transfer in this case leads to an eventual disruption of the donor on dynamical timescales. 
Dynamically important nuclear burning may occur during the merger process \citep{Metzger2012}. The event may be the origin of calcium-rich gap transients. \citet{Lyman+2013} list twelve such events. These transients have luminosities between those of novae and supernovae and durations of about a month, relatively short compared to type II supernovae. 

In order to better determine the dividing line between stable and unstable mass transfer we revisit the standard model of mass and angular momentum loss in WD--NS binaries.  We model mass transfer using a modified smoothed particle hydrodynamics technique. This allows us to measure the amount of angular momentum lost by the binaries through disc winds, and also to measure the effect of binary eccentricity on mass transfer. We then use our results to construct a model of the long-term evolution and observational appearance of these binaries after they come into contact.

The standard model of mass loss in WD--NS binaries assumes that the material is lost entirely through a jet with the specific angular momentum of the accretor \citep{Tauris1999}. However, the majority of sources reach significantly super-Eddington  accretion rates during the early stages of their evolution \citep{2012A&A...537A.104V}. At these rates disc winds become important, which is evidenced by the change in disc structure seen in simulations \citep{Scadowski2014}, as well as by observations of rapidly accreting systems such as SS~433 \citep{Begelman2006}. Disc winds are expected to 
remove more angular momentum than mass loss in a jet.  Our models allow us to measure the amount of angular momentum lost in the disc winds and hence probe the stability of the mass transfer process.

Our hydrodynamic modelling continues the work done by \citet{Church2009} and \citet{Lajoie2011a}, both of whom introduced a method to model realistically low mass-transfer rates. Our numerical method improves on the technique proposed by \citet{Church2009}, whereas the idea of using the results of hydrodynamic simulations for long-term modelling is inspired by the first thorough study of mass transfer in eccentric binaries \citep{Lajoie2011}. 
Our long-term evolution model for WD--NS binaries  follows earlier work by 
 \citet{Sepinsky2009} and \citet{Dosopoulou2016a, Dosopoulou2016}, who focused specifically on mass-transfer
 in eccentric systems, and is complementary to the work by \citet{2012A&A...537A.104V}, as it accounts for the angular momentum loss through disc winds and for the evolution of unstable systems.

The remainder of this paper is structured as follows.  In Section~\ref{PopObs} we summarize what is known observationally and from population synthesis calculations about WD--NS binaries before they come into contact. In Section~\ref{sec:Num} we explain our hydrodynamic model; we verify its applicability to circular and eccentric WD--NS binaries in Section~\ref{sec:ResMain}.  Section~\ref{sec:Flows} contains our analysis of the mass flows that our simulations exhibit, which we use in Section~\ref{sec:DiscussionMain}
to model the long-term evolution and observability of post-contact WD--NS binaries. We summarize our conclusions in Section~\ref{sec:ConclusionsMain}.

\section{Populations and observed systems}
\label{PopObs}

Before a WD--NS binary comes into contact the neutron star may potentially be observed as a radio pulsar. About $150$ pulsars are observed to have a companion that is most likely a white dwarf of mass above $0.08\,M_\odot$ \citep[ATNF pulsar catalogue,][]{Manchester2005}.  Ten of these are expected to come into contact in less than a Hubble time. Theoretically, the field population of WD--NS binaries may be divided according to how we believe that they formed.  Additionally, there are two separate dynamical formation channels that are only active in globular clusters.

\subsection{Field populations}

We divide the field population of WD--NS binaries into two groups, depending on whether the white dwarf formed before the neutron star (field population one) or after the neutron star (field population two).  

In field population one neither star initially has enough mass to produce a neutron star via a supernova explosion. However, mass transfer from the primary to the secondary (when the primary evolves) can be sufficient to increase the secondary's mass to the point where it can produce a neutron star.
For this scenario to work the initial mass ratio of the binary must be close to unity; hence field population one is predicted to contain binaries with a massive WD. As the NS is formed second, the newly formed binaries are eccentric ($e\approx 1$) due to the NS kick. PSR
J1141-6545 \citep{Kaspi2000} is the only known population one WD--NS binary that will merge in less than a Hubble time. It has an orbital period of $4.74\,\textrm{h}$, an eccentricity of $0.17$, and a lower bound on the WD mass of $0.98\,M_{\odot}$ \citep{Bhat2008}. A detailed analysis of the progenitor evolution scenarios for this population may be found in \citet{Church2006}. 

In population two, the neutron star is formed first, from the more massive star. In order for the original binary to survive the neutron star kick, an episode of common-envelope evolution is necessary to shrink the orbit before the supernova occurs. As the binary is sufficiently tight at this stage, it typically comes into contact when the secondary becomes a red giant, before helium ignition. Therefore, population two typically leads to low-mass helium WDs in a binary with a neutron star. Higher mass WDs are also possible, in which case the NS is likely to be more massive \citep{Antoniadis2013}. The orbit is circular due to mass transfer taking place between the supernova and the formation of the white dwarf. Five circular field binaries are known which contain a WD of mass above $0.08\,M_\odot$ and merge in less than a Hubble time (ATNF catalogue). Of these, three binaries contain He WDs, one contains a CO WD, and one an ONe WD. Detailed evolutionary scenarios for this population may be found in \citet{Tauris2011} and \citet{Antoniadis2013a}.

Populations one and two are expected to have comparable coalescence rates \citep{Davies2002,Nelemans2001}. There is a bias towards observing members of population two, since the neutron star in population one forms after all the episodes of mass transfer and hence cannot be recycled. Since non-recycled pulsars have 
short spin-down timescales most WD--NS binaries in population one are radio-quiet. 

\subsection{Populations in globular clusters}

Another important site for WD--NS binary formation is the cores of globular clusters.
Even though globular clusters contain only about $0.1$ per cent of the stars in the
Galaxy, they contain about a third of the observed WD--NS systems (ATNF
catalogue) and about a half of all known UCXBs \citep{2012A&A...543A.121V}.
Additional channels of WD--NS formation must therefore be present. We outline two such possible channels (populations three and four).

Population three is proposed to form when free neutron stars collide with or are
tidally captured by red giants \citep{Verbunt1987}. The red giant's envelope is
subsequently ejected through a common envelope phase and a tight, eccentric
WD--NS binary with a low-mass companion is produced \citep{2005ApJ...621L.109I, 2006ApJ...640..441L}. This
scenario may explain the population of UCXBs in globular clusters
\citep{2008MNRAS.386..553I}.

Since the number of UCXBs observed in globular clusters is comparable to the number known in the field, globular clusters are expected to harbour a comparable number of WD--NS binaries which will come into contact in less than a Hubble time. Yet, presently, there are no detached eccentric WD--NS systems observed in globular clusters which will merge in less than a Hubble time (ATNF catalogue). The binaries in population three form relatively tight \citep{2005ApJ...621L.109I}. Since the gravitational wave inspiral time is a strong function of $a$ \citep{Peters1964}, their lifetime is expected to be short. This may explain the lack of observed detached systems, as well as the young age of the persistent UCXBs in globular clusters \citep{2005ApJ...621L.109I}. Another possible explanation would be that the neutron stars involved in this scenario are typically non-recycled. The eccentricity may have been indirectly observed in $4$ UCXBs, all
of which are located in globular clusters \citep{2007MNRAS.377.1006Z,
Prodan2015}. It is thought, however, to be induced and be varying due to a third companion, which leads to long-term modulations in the observed {\it X}-ray luminosity.

Population four is produced in exchange interactions in globular clusters. In this scenario the WD--NS binary forms as a result of an encounter between a single NS and a binary containing white dwarfs and/or main sequence stars \citep{Davies1995, 2008MNRAS.386..553I}. Since low-mass companions are preferentially ejected in triple encounters, binaries in this population are expected to contain higher-mass WDs and may therefore be a source of gap transients in globular clusters. At present there are no observed detached binaries thought to belong to this population. This may be explained by the neutron stars  remaining non-recycled and thus radio-quiet.

\subsection{State upon coming into contact}

\begin{figure}
\includegraphics[width=\columnwidth]{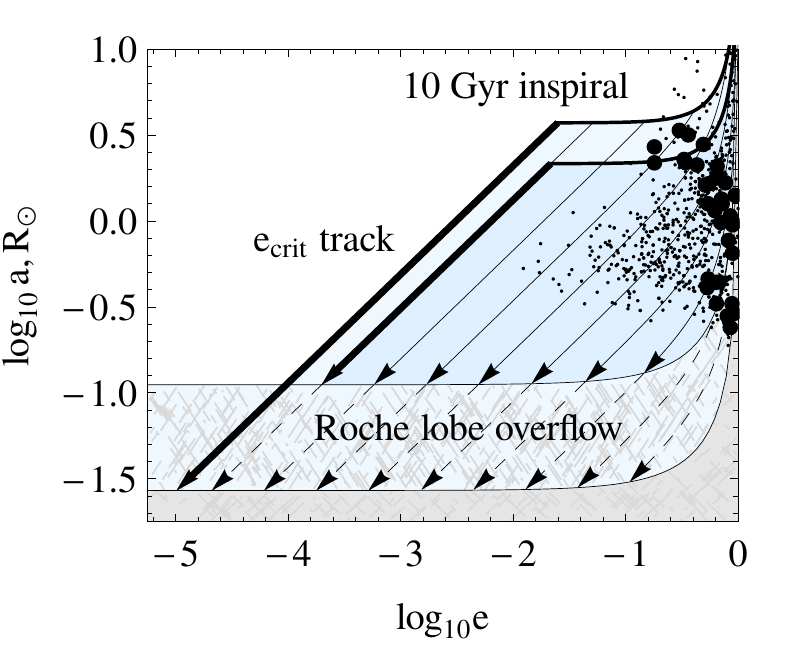}
\caption{Evolutionary tracks showing the orbital semi-major axis $a$ as a function of orbital eccentricity $e$ for WD-NS binaries undergoing gravitational wave emission. The large black points represent simulated systems from  \citet{2006ApJ...640..441L} produced through dynamical collisions of neutron stars and red giants in globular clusters. The small black points represent eccentric WD--NS binaries of field population one  from the population synthesis model of \citet{Church2006}. The blue shaded regions correspond to systems whose eccentricities are large enough that they experience significant variations in the mass-transfer rate across a single orbital period when they come into contact. The larger pale-blue shaded region of the diagram is for the high-mass white dwarf systems ($M_{\rm WD}=0.9\,M_\odot$) and the darker blue region is for low-mass white dwarf systems ($M_{\rm WD}=0.15\,M_\odot$). The cross-hatched area shows where white dwarfs overflow their Roche lobes, upper and lower edges for low- and high-mass white dwarfs.}
\label{fig:GCsPopEcc}
\end{figure}

We have evolved the orbits of the synthetic population one \citep[from][]{Church2006} and the simulated systems for population three \citep[from][]{2006ApJ...640..441L} forward in time until contact using the equations of \citet{Peters1964}:
\begin{equation}
\left(\dfrac{\dd a}{\dd t}\right)_{\rm GW} = - \dfrac{64}{5} \dfrac{G^3 M_1 M_2 (M_1 + M_2)}{c^5 a^3 (1-e^2)^{7/2}}\left(1+\dfrac{73}{24}e^2+\dfrac{37}{96}e^4\right)
\end{equation}
\[
\left(\dfrac{\dd e}{\dd t}\right)_{\rm GW} = - \dfrac{304}{15} e  \dfrac{G^3 M_1 M_2 (M_1 + M_2)}{c^5 a^4 (1-e^2)^{5/2}}\left(1+\dfrac{121}{304}e^2\right)
\]
These equations describe the evolution of the semimajor axis $a$ and eccentricity $e$ of a binary with masses $M_1$, $M_2$ due to the emission of gravitational waves, assuming that this is the only process affecting the orbital evolution. The results are shown in Figure~\ref{fig:GCsPopEcc}.  We compare the eccentricities at contact to the critical eccentricity needed for the mass transfer to turn on and off during the orbit, $e_{\rm crit}$.  This occurs when the amplitude of variation of the Roche lobe radius during the orbit is larger than the scale height $h_{\rho}$ of the white dwarf's atmosphere.  Hence
\begin{equation}
e_{\rm crit}=h_{\rho}/R_{\rm WD}=\frac{R_{\rm WD}}{GM_{\rm WD}}\frac{\mathcal{R}T_{\rm eff}}{\mu},
\label{eqn:ecrit}
\end{equation}
where we have assumed an ideal gas atmosphere: see Section~\ref{sec:SingleSetup} for a more detailed discussion of this assumption.  We find that the eccentricity at contact is $10$ to $1000$ times larger than $e_{\rm crit}$ for both populations. For example, for population one the eccentricities at contact are between $10^{-3.5}$ and $10^{-2}$, whilst the density scale height in the outer parts of the white dwarf atmosphere $h_{\rho}$ should be  $10^{-5}$ to $10^{-6}\,R_{\rm WD}$.  Hence it is likely that initially mass transfer will turn on and off during each orbit of the binary.

This calculation does not take into account tidal interactions, which circularise the binaries.  However, we expect that tides are unlikely to reduce $e$ by two to three orders of magnitude before mass transfer starts.  \citet{Fuller2012-2} find that double white-dwarf binaries only reach spin synchronisation $\delta \omega/\omega$ of $10^{-1}$~--~$10^{-2}$ by contact time, where $\omega$ is the orbital frequency of the binary and $\delta \omega$ is the difference between $\omega$ and the spin frequency of the white dwarf. In general spin synchronisation is faster than orbital circularisation \citep{Zahn1977, Hut1981} since the stars have smaller moments of inertia than the orbit and hence less angular momentum is required. WD--NS binaries, therefore, will likely not have enough time to circularise before the
onset of mass transfer.

\begin{figure}
\includegraphics[width=\columnwidth]{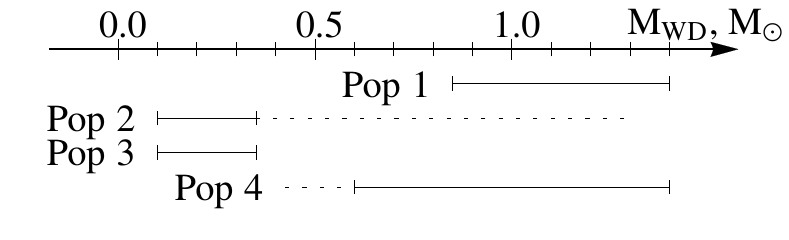}
\caption
{
Mass distributions of the populations of white dwarf neutron star binaries. The field population of eccentric binaries where the white dwarf formed first (population one) is expected to always contain high-mass white dwarfs. The field population of binaries where the neutron star formed first (population two) contains mostly low-mass helium white dwarfs. Population three, produced by dynamical collisions in globular clusters, is expected to contain only low-mass helium white dwarfs. Population four, produced by dynamical exchange interactions in globular clusters, is expected to contain mostly high-mass white dwarfs. Our results imply that populations two and three contain most of the progenitors of UCXBs, whereas populations one and four contain most of the binaries which lead to transient outcomes.
}
\label{fig:PopulationsChannels}
\end{figure}

We summarise the four populations of WD--NS binaries in Figure~\ref{fig:PopulationsChannels}. Each population is characterised by the range of  white-dwarf masses the binaries contain. 
Hence, by determining the critical white-dwarf mass separating stable and unstable mass transfer, we will be able
to determine which populations will contribute to the population of ultra-compact {\it X}-ray binaries (UCXBs,
produced via stable mass transfer) and which will contribute to unstable systems (possibly
producing other, visible, transient sources).

\section{Numerical methods}
\label{sec:Num}

\subsection{Modified SPH code details}
\label{SPHOWat}
We model the hydrodynamics of mass transfer using a modified SPH method. We
incorporate recent developments in SPH, which are
discussed in e.g. \citet{Rosswog2014}.  A general introduction to SPH may
be found in a number of reviews
\citep{Monaghan2005,Rosswog2009,Springel2010,Price2012}.

Due to its Lagrangian nature, SPH is a standard method of choice for modelling compact binary mergers. It is challenging to use, however, when it comes to modelling the onset of mass transfer or stable mass transfer in real UCXBs. Resolving realistically low mass-transfer rates would require the use of at least $10^{12}$ particles. For example, the well studied UCXB system 4U~1820-30 transfers about $3\times 10^{-12}\,M_{\rm WD}$ of material per orbit \citep{Stella1987}. The issue is particularly important for eccentric binaries, where the mass-transfer rate varies over the orbit.

The earliest hydrodynamical study of eccentric systems by \citet{Regos2005} used standard SPH to test the applicability of the Roche lobe formalism. \citet{Church2009} and \citet{Lajoie2011a} each introduced a new method to model realistically low mass-transfer rates in eccentric systems. The study by \citet{Church2009} allowed for a flexible boundary for the atmosphere and arbitrarily low mass-transfer rates. \citet{Lajoie2011a} used a fixed boundary based on ghost particles, which reduced the computational cost, and also accounted for the extended size of the accretor. The accompanying paper \citet{Lajoie2011} presented the first thorough study of mass transfer in eccentric binary main-sequence stars and investigation of the orbital mass-transfer profile, mass loss and effects of mass transfer on stellar structure.

We make use of the Oil-on-Water SPH modification, proposed in \citet{Church2009}. The key idea for the method is to represent the stellar body and the atmosphere by separate types of SPH particles (Water and Oil, respectively). Having them separated by carefully chosen artificial forces (Appendix~\ref{SPHOWForce}), we assign the Oil particles arbitrarily low masses and hence model realistically low mass-transfer rates. The Oil-on-Water method allows us to steadily transfer large numbers of Oil particles without affecting the Water body of the star. In the standard SPH approach transferring a large number of particles from the stellar body implies that the donor is tidally disrupted.

We use an SPH formulation derived from a Lagrangian \citep[as opposed to a `vanilla'
formulation, see e.g.][]{Rosswog2009}. This provides better energy
conservation when the SPH smoothing length $h_i$ changes with time, as is the case
for the mass-transfer process. We use  Wedland W6 kernels \citep{2012MNRAS.425.1068D}, which prevent the occurrence of the pairing
instability by making it energetically unfavourable for SPH particles to spontaneously merge and form pairs. Furthermore, they allow us
to use arbitrary numbers of SPH neighbours ($400$ in our case), which reduces
the so-called E0 error, a discretisation error which acts as a spurious numerical force. We use the same particle mass $m_i$ for all the Water particles, because varying $m_i$ has been reported to lead to numerical
artefacts \citep{Dan2011, LorenAguilar2009}. Similarly, we use a different constant value for the masses of the Oil particles, which are kept spatially separated from the Water particles.

Our code is based on our own $\rho-h$ relation of the form $(\rho+\rho_0)h^3=\eta m$, following the idea by \citet{Monaghan2005} (their Equation~4.5) as summarized in Appendix~\ref{SPHRhoH}. The method allows us to set a maximal value for $h$ in a self-consistent conservative way (we use $h_{\rm max}=R_{\rm WD}$). It also partially mitigates the so-called fallback problem, which occurs when particles that have been ejected to large distances  subsequently fall back on to the star.  Because the ejected particles acquire large smoothing lengths during their excursion through low-density regions, when they fall back on to the surface of the donor star they interact with very many particles, which is computationally expensive. The artificial conductivity is implemented as in \citet{Hu2014}. However, we control it by the artificial viscosity parameter $\alpha$ (instead of $\xi^i$ in their Equation~23). This way, artificial conductivity doesn't trigger at small velocities and appears only near shocks. We use the gravitational tree from \citet{Benz1990}. The integrator is based on the symplectic KDK scheme from \citet{2010MNRAS.408..669C}, modified in a time-symmetrical way to individual timesteps, following  \citet{1997astro.ph.10043Q} and \citet{2005MNRAS.364.1105S}. The timestep limiter is based on the scheme by \citet{2009ApJ...697L..99S}. It prevents particles from having neighbours with significantly different  timesteps. This condition is necessary for correct treatment of shocks. We also ensure that none of $h$, $\rho$ or $u$ change by more than five per cent over a single timestep.

We derive the expressions for the boundary Oil-Water forces by starting from a Lagrangian (Appendix~\ref{SPHOWForce}). This ensures that the forces are curl-free and thus removes spurious atmosphere heating. By choosing a suitable interaction potential we ensure that no mixing takes place between the Oil and Water phases. We additionally ensure that the boundary forces only extend up to half-way
through the atmosphere. Therefore the particles participating in mass transfer
are purely pressure supported. We reach a linear computational complexity in
$N_{\rm oil}$ for the gravity force, which provides us with a significant
performance benefit. This is possible because the white dwarf atmospheres contain
sufficiently little mass that they essentially do not self-gravitate. In
practice we utilise two separate trees, for Water and Oil particles.

We use a modified version of the Helmholtz equation of state for the Water particles \citep{Timmes2000}. We include perfect degenerate and ideal gas (electrons and ions) components. Particle compositions are kept constant. We do not include Coulomb corrections because they are not known for the whole $(\rho, T)$ domain of interest and may lead to discontinuities in the $u(T)$ dependence.

We omit radiation pressure, because using it is not physically self-consistent without accounting for radiation transfer. These simplifications are appropriate for the WD interior, which is almost isothermal \citep{Mestel1967}. 
The atmospheres of white dwarfs which are Roche-lobe filling are likely substantially more complex, with tides and shocks from the mass-transfer flows heating the material on the one hand, and radiative cooling on the other.  To a first approximation, tidal heating dominates the energy input to the outer layers, and leads to typical equilibrium temperatures of $3-5\times10^4\,\textrm{K}$ \citep{Fuller2013}.
We use the mass-transfer model of \citet{Ge2010} (discussed further is Section~\ref{sect:CircularVerification}), in conjunction with our equation of state, to identify which atmospheric layers contribute to the transferred mass. For typical mass-transfer rates of less than about $10^{-5}\,M_\odot/\textrm{yr}$ material is transferred from the outermost layers, which behave as an ideal gas. Hence we use an ideal gas equation of state for the Oil particles. This has an additional benefit of making the SPH equations scale-free in the mass of Oil particles. To avoid having to model the complex thermal energy flows in the envelope, which are beyond the scope of our approach, we maintain the envelope at an equilibrium temperature.  This temperature is chosen to make the mass transfer representative of that from white dwarfs in thermal equilibrium from tidal heating, as discussed in Section~\ref{sect:connection}.

Accounting for ionisation may possibly be important in modelling mass transfer from white dwarf donors, in particular the ejected material \citep{2013A&ARv..21...59I}. The Helmholtz equation of state assumes complete ionisation. However, complete ionisation temperatures for He, C, O and Ne are $6\times 10^5\,\textrm{K}$, $6\times 10^6\,\textrm{K}$, $1\times 10^7\,\textrm{K}$ and $1.5\times 10^7\,\textrm{K}$. These temperatures are much higher than those on the WD's surface and  much lower than those measured for the ejecta, as shown in Section~\ref{sec:Flows}. The ionisation energy budget, therefore, may have an effect on the interaction between the newly transferred material and the disc, as well as on the cooling of the ejected material.

We test the reliability of our code by measuring the fractional drifts in total
$J_z$ and $E$. They remain below $10^{-5}$ and $10^{-4}$, correspondingly, over
tens of orbital periods.  The major contributors to the drift are the
gravitational tree and the integrator.

\subsection{Single stellar model setup}

\label{sec:SingleSetup}

In our simulations we focus on four models of white dwarfs:
\begin{itemize}
\item A purely He model
with $M_{\rm WD}=0.15\,M_\odot$, which is a typical UCXB progenitor,
\item A $0.6\,M_\odot$
CO model with $X_{\rm C}=X_{\rm O}=0.5$, a typical CO white dwarf,
\item A $1.0\,M_\odot$ O-Ne model with
$X_{\rm O}=0.75$, $X_{\rm Ne}=0.25$, a typical heavier white dwarf, and
\item A $1.3\,M_\odot$ O-Ne model with $X_{\rm O}=0.75$, $X_{\rm Ne}=0.25$, the WD
type expected to produce the brightest transients.
\end{itemize}
The neutron star companion is represented as a point-like object with a mass of $1.4\,M_\odot$. We soften the gravity of the neutron star by using a Wedland W6 kernel with a core radius of $0.2\,R_{\rm WD}$, the exact value having no effect on the measured quantities. The SPH models of the white dwarfs contain $10^5$ Water and $3\times 10^5$ Oil particles. The Water particles representing the degenerate stellar body are
set to have a constant $T=10^5\,\textrm{K}$, whilst the ideal gas atmosphere represented by the Oil particles is set to be
fully ionised and initially isothermal. Theoretical stellar profiles for the SPH models are obtained by writing the standard stellar structure equations \citep{Kippenhahn2012} in $\ln \rho$ coordinates and solving them under the Helmholtz equation of state by using our own code.

We implement a new method for setting up a single SPH star. The star is set up in spherical shells, one by one, starting from the centre. The particles on each shell are arranged equidistantly in a spiral pattern, following \citet{saff1997distributing}. Each shell is defined this way by its radius $r_i$ and mean inter-particle distance $l_i$, defined by
\begin{equation}
\rho(r_i)l^3_i = \dfrac{9\sqrt{3}}{4\pi} \dfrac{M_{\rm WD}}{N_{\rm total}},
\end{equation}
where $\rho(r)$ is provided by the theoretical stellar profile and $N_{\rm total}$ is the desired number of SPH particles in the star. The next shell has $r_{i+1}=r_i + l_i$ and $l_{i+1}$ is defined by the above equation. The method provides a locally isotropic particle order, i.e. the mean particle packing densities in radial and tangential directions are nearly equal. This ensures that the newly set up SPH particles have no preferred direction to move, i.e. they are close to equilibrium. The Oil atmosphere is set up following a similar strategy. A similar approach based on the spiral patterns on a sphere has been recently introduced by \citet{Raskin2016}.

In order to follow flows in the Oil layer reasonably,  the atmospheric scale height, $h_{\rho}$, in our SPH models must be larger than the typical size of Oil particles, $h_{\rm SPH}$.  To achieve this we adjust the temperature of the Oil envelope: 
since Oil particles follow the ideal gas equation of state, $h_\rho$ is proportional to their temperatures. 
We require that $h_{\rho}\approx0.03\,R_{\rm WD}$, which corresponds to temperatures of above $10^6\,\textrm{K}$.
Real white dwarfs have much smaller scale heights $h_\rho$ of $10^{-3}-10^{-5.5}\,R_{\rm WD}$, and correspondingly cooler atmospheres. 
We can still describe mass transfer from real white dwarfs using our
models, despite the large difference in the atmospheric depth, if the mass-transfer rate depends only
on the number of scale heights by which the Roche lobe digs into the atmosphere at pericentre.  We use our simulations to test this key assumption and discuss it further in Section~\ref{sec:ResMain}.

An additional effect of choosing the atmospheric scale height to be comparable to the SPH smoothing radius is that the actual scale height of our models varies somewhat from that predicted by models of isothermal stellar atmospheres.  We 
measure the scale height $h_\rho$ of our models from the gradient of their density profile after they have been relaxed and find that it is about 
$30$ per cent larger than expected. We have verified analytically that this factor is expected from the SPH equations and is due to the particle resolution in the atmosphere being comparable to $h_\rho$. The limited resolution also leads to slow shock heating, such that the internal energy of the Oil particles increases by about $20$ per cent over $20$ orbital periods. We avoid this contaminating our results by adding a cooling term in the binary runs which adiabatically restores the atmospheric temperature of the white dwarf to its original value on the timescale of half an orbital period.

\subsection{Binary setup method}
\label{sec:BinSetup}

We relax the binary using a scheme in which the final relaxed state corresponds to ongoing mass transfer. The relaxation is done in stages. Initially the Water star is set up in isolation and relaxed under a viscous damping force over five dynamical times. An Oil atmosphere is added and the system is relaxed again. The white dwarf is then set in a binary with a neutron star at a separation of $1.8\,a_{\rm RLO}$, where $a_{\rm RLO}$ is the semi-major axis where the real white dwarf would just fill its Roche lobe.  Relaxation is performed in a corotating frame over two binary periods. During this phase we linearly decrease the orbital separation with time till the binary is at the pericentre of its desired final orbit. Mild viscously damped mass transfer starts during the inspiral process, and we remove particles that reach half-way between $L_1$ and the neutron star in order to ensure that no particles fall directly onto the neutron star. Finally, we let the proper oscillations of the white dwarf dampen over one period in a corotating frame with the binary set at pericentre, mass transfer ongoing and the damping set on. Since the tidal forces are strongest at the pericentre, the setup is expected to correspond well to real systems and to have a low level of spurious proper oscillations.

\begin{table}
\caption{The effect of quadrupole interactions on binary orbits.}
\begin{center}
\begin{tabular}{|c|c|c|c|c|}
\hline 
$M_{\rm WD},\,M_\odot$ & $0.15$ & $0.6$ & $1.0$ & $1.3$ \\ 
\hline 
$\delta e_{\rm quadr}$ & $4.0\cdot 10^{-3}$ & $7.0\cdot 10^{-3}$ & $6.8\cdot 10^{-3}$ & $7.0\cdot 10^{-3}$ \\ 
$A_q$ & $6.02$ & $4.90$& $3.72$& $3.34$ \\ 
$B_q$ & $8.88$ & $8.70$& $6.64$& $6.96$ \\ 
\hline 
\end{tabular}
\end{center}
Notes: $\delta e_{\rm quadr}$ is the effective eccentricity introduced if
quadrupole interactions are not accounted for. The corresponding correction to
the angular velocity $\omega$ is given by  $\delta e_{\rm quadr} = 2\delta
\omega/\omega_{\rm Kep}$.  $A_q$ and $B_q$ are coefficients in our linear fit to the radial-radial quadrupole moment, as given in Equation~\ref{eqn:Dquad}.
\label{table:Quadrupole}
\end{table}

We find that accounting for the  quadrupole moment of the mass distribution in the white dwarf is necessary in order to 
accurately calculate the gravitational force between the two stars in the binary, and thus
calculate correctly the orbital period. The physical size of the white dwarf is comparable to that of the orbit, so the tidal force from the neutron star deforms it such that it becomes aspherical. Hence the potential for gravitational interaction in the binary is not perfectly represented by a point mass approximation, i.e. is not Keplerian. Setting a binary on an orbit assuming a Keplerian potential produces a spurious $e \approx 0.01$, which in turn produces spurious oscillations in the mass-transfer rate.  We measure the radial-radial quadrupole moment for each of our four white dwarf models over a range of relevant separations 
($1.15 < a/a_{\rm RLO} < 1.8$), and find that it can be fit as
\begin{equation}
D_{\rm rr}(a)=10^{-2}M_{\rm WD} R_{\rm WD}^2(A_q-B_q(a/a_{\rm RLO}-1)),
\label{eqn:Dquad}
\end{equation}
where values for the constants $A_q$ and $B_q$ are given for each white dwarf model in Table~\ref{table:Quadrupole}.

To set the binaries up we first obtain the empirical interaction potential by measuring the gravitational force as a function of binary separation during the inspiral phase. This allows us to set the binary on any desired circular orbit. The same interaction potential is used to set the binaries on eccentric orbits in the eccentric runs.  We find that this empirically derived gravitational potential is consistent with one derived by including the quadrupole moments of the mass distribution; this implies that the tidal distortion of the white dwarf is responsible for the deviation from a Keplerian orbit. By using this method we reduce the spurious induced eccentricity by about an order of magnitude, to $\delta e\approx 5\times 10^{-4}$. This provides a smoother $\dot{M}$ profile, e.g. see Section~\ref{EccMTGen}, which is necessary in order to set up a circular binary. We argue that this effect explains the oscillations observed in Figure~12 of \citet{Dan2011}.

Though important for binary setup, it appears that the quadrupole interaction has
little effect on mass transfer in real binaries. With $a$ and $e$ fixed
the change in the interaction potential mainly affects the orbital period,
with the typical change being $0.1$--$1$ per cent. It may be important, however,
for the phasing of gravitational wave profiles of WD--NS binaries when they come
into contact \citep{Sravan2014}.

Another effect one has to take care of when preparing initial conditions is the change of stellar radius of the donor with binary separation.  \citet{Dan2011} found that mass transfer in their simulations started at separations $14$ per cent larger than expected from the formula of \citet{Eggleton1983}. We observe a similar effect in our simulations, where the Oil layer extends to $1.15\,R_{\rm WD}$ and mass transfer starts at $\sim 1.3\,a_{\rm RLO}$. We argue that the early mass-transfer onset takes place in part due to the volume change of the tidally deformed donor star.  We measure the volumetric radius of the SPH star by Delaunay tessellation \citep{Delaunay1934}, accounting only for the particles which have not yet crossed the $L_1$ plane. The tessellation corresponds to constructing a convex hull with the vertices located at the positions of SPH particles. We measure the relative radius change $\delta R_{\rm WD}/R_{\rm WD,\infty}$ to grow from $6$ to $8.5$ per cent as the binary separation decreases from $1.3\,a_{\rm RLO,Eggl}$ to $1.15\,a_{\rm RLO,Eggl}$.  The $1.3\,M_\odot$ model exhibits a stronger deformation, with $\delta R_{\rm WD}/R_{\rm WD,\infty}$ growing from $12$ to $17$ per cent over the same interval. Accounting for the change in $R_{\rm WD}$ is necessary to accurately measure the $\dot{M}(\Delta R)$ dependency (see Section~\ref{sect:CircularVerification}).

\begin{figure}
\includegraphics[width=\columnwidth]{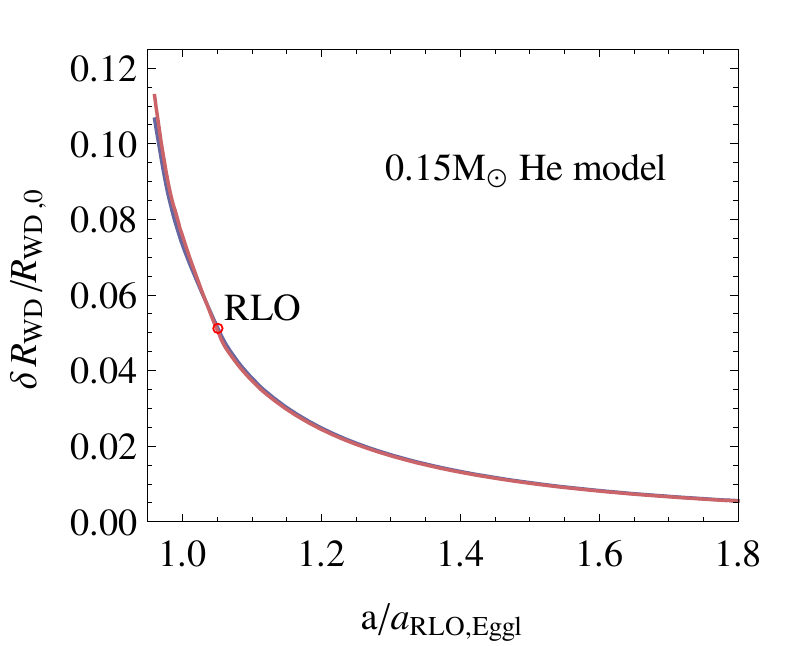}
\caption{
The fractional increase in donor radius due to tidal deformation, $\delta R_{\rm WD}/R_{\rm WD}$, as a function of binary separation $a$. The curves are obtained by modelling a $0.15\;M_\odot$ He WD at a set of different resolutions of up to $8\times 10^5$ SPH particles. For each value of $a$, $\delta R_{\rm WD}$ was estimated by fitting $\delta R_{\rm WD,i}$ at different resolutions (i.e. number of SPH particles, $N$) as a linear function of average particle separation ($N^{-1/3}$) and extrapolating to $N^{-1/3}\rightarrow 0$. Water-only models were used so as to make sure that the effect is generic and not caused by our Oil-on-Water modification. The volumetric radius change was measured by using an SPH volume estimate (blue curve) and by Delaunay tessellation of a convex hull spanned by the SPH particles (red curve). Mass transfer starts at a separation about 5 per cent larger than derived by using isolated stellar models.
}
\label{fig:DeltaRPlot}
\end{figure}

Furthermore, we verified that the measured change in $R_{\rm WD}$ converges with increasing resolution and that is not an effect of the tessellation scheme or the Oil-on-Water method. We
constructed a set of $0.15\,M_\odot$ Water-only models at resolutions of
$(50,100,200,400,800)\times 10^3$ particles, and measured their radii between $1.8\,a_{\rm RLO}$ and $0.95\,a_{\rm RLO}$ by two methods: tessellation (as before) and
a standard SPH estimate $V=\sum_i m_i/\rho_i$. The latter method is unavailable
for the Oil-on-Water case because of the artificial gap between the Oil and Water. We
estimated $\delta R_{\rm WD} (a)$ by fitting a straight line to $R_{{\rm WD}, i}$
as a function of $N^{-1/3}$ for models at different resolutions (i.e. number of SPH particles, $N$) and then 
extrapolating to $N^{-1/3}\rightarrow 0$.  
The results are shown in
Figure~\ref{fig:DeltaRPlot}. We find that the separation at Roche-lobe overflow exceeds the Eggleton estimate by $5.1$ per cent. 
The relative error in our measurement of $\delta R_{\rm WD} (a)$ 
induced by accounting for the finite resolution of our models is less than $0.04$ at a $0.95$ confidence level.  
The error induced by the choice of volume measurement technique is smaller still.

%We find that the separation at Roche-lobe overflow exceeds the Eggleton estimate by $5.05$ or $5.11$ per cent depending on the radius estimate used, whilst the relative error for $\delta R_{\rm WD} (a)$ is less than $0.040$ and $0.033$ at a $0.95$ confidence level.

\section{SPH simulations: Method verification}

\label{sec:ResMain}

We first apply our method to circular systems, where the mass-transfer rate is well-known from analytic studies, to verify that our method is valid.  We then consider eccentric systems, and test that our method can be scaled to model real WD--NS binaries.  Finally we analyse the properties of the accretion flows that we observe in our models, and extract properties of them that we can use for longer-timescale simulations.

\subsection{Verification of the Oil-on-Water method using circular systems}

\label{sect:CircularVerification}

Since the outer atmosphere of a white dwarf is an ideal gas, the instantaneous mass transfer rate in a circular WD--NS binary should follow the model of \citet{Ritter1988}. He parametrises the mass-transfer rate as a function of the degree of Roche-lobe overflow,
\begin{equation}
\Delta R= R_{\rm WD} - R_{\rm RL,WD}.
\end{equation}
as
\begin{equation}
\label{MDotInst}
\dot{M}_{\rm circ}=\dot{M}_0 \exp\left(\dfrac{\Delta R}{h_\rho}\right),
\end{equation}
where $h_\rho$ is the density scale height of the white dwarf's atmosphere. We test this by placing each model at a separation of $1.3\,a_{\rm RLO}$ and letting it freely transfer mass. We then decrease the binary separation adiabatically to $1.15\,a_{\rm
RLO}$, whilst preserving corotation. The procedure was repeated for the inspiral
times of $0.5$, $1$, $2$ and $4$ binary periods. 
For this test we remove the SPH particles half-way between the $L_1$ point and the neutron star in order to simplify the measurement of how much mass has been transferred and reduce the computation time. 

\begin{figure}
\includegraphics[width=\columnwidth]{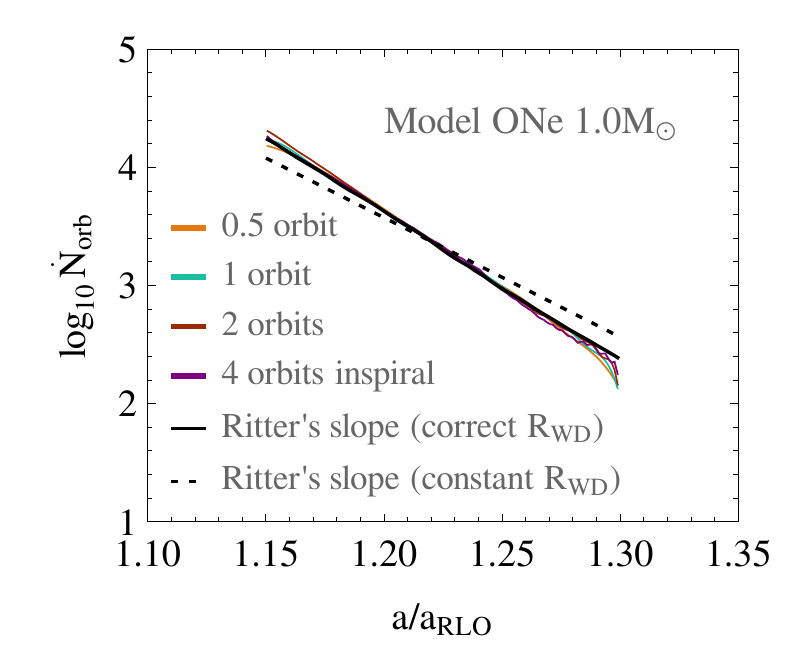}
\caption{
The instantaneous mass-transfer rate, given as the rate at which particles are
transferred from the white dwarf, $\dot N_{\rm orb}$, is plotted as a function
of separation $a$. The curves are obtained by adiabatically contracting the
binary orbits over half an orbital period (orange line), one orbital period
(green line), two orbital periods (brown line) and four orbital periods
(purple line).  For comparison we plot the slope from \citet{Ritter1988} using a
constant white-dwarf radius (dashed black line) and one that varies with the
tidal distortion of the white dwarf (solid black line).  Our simulations
reproduce Ritter's slope remarkably well.  The feature in the beginning of
inspiral is due to initial conditions slightly inhibiting the initial mass
transfer.
}
\label{fig:InstMTVersusA}
\end{figure}

The results are shown in Figure~\ref{fig:InstMTVersusA}.
The observed and expected dependencies match very well up to the constant factor $\dot{M}_0$; the observed difference is between $5$ and $10$ per cent, depending on the white dwarf model and once the variation of white-dwarf radius with separation has been taken into account. The fact that the $\dot{M}$ dependence is the same for an inspiral over $0.5$ orbital periods implies that the donor can adjust on timescales shorter than the orbital period.

The pre-factor $\dot{M}_0$ in Equation~(\ref{MDotInst}) may be obtained from the analysis of \citet{Ge2010}, which is the latest revision of a mass loss law for donors with arbitrary equations of state. By comparing it to the values from our simulations we observe a difference of a factor of $11$ to $14$. In other words, the simulated stars transfer the shell of material outside the Roche surface about ten times faster than expected from analytic modelling.  Given the approximate character of the pre-factor in the analytic formula \citep{Ritter1988,Ge2010}, we find it more likely to be the reason for the discrepancy. If the discrepancy is an SPH effect, it may be due to the Oil atmospheres having much larger scale height than real stars. 
In any case, the exact value of the pre-factor $\dot{M}_0$ has no effect on long-term evolution \citep{Webbink1985}.  This is because the exponential dependence on separation in the mass-loss law means that a large change in the $\dot{M}_0$ is equivalent to a very small change in $a$.  For a binary spiralling in from large separations under the effect of gravitational radiation, a making $\dot{M}_0$ larger by, say, an order of magnitude, simply means that a given mass-transfer rate is attained at a separation that is wider by a few density scale heights. This is a negligible fraction of the binary separation and hence makes no difference to the orbital evolution.

\begin{figure}
\begin{minipage}{\columnwidth}
\includegraphics[width=\columnwidth]{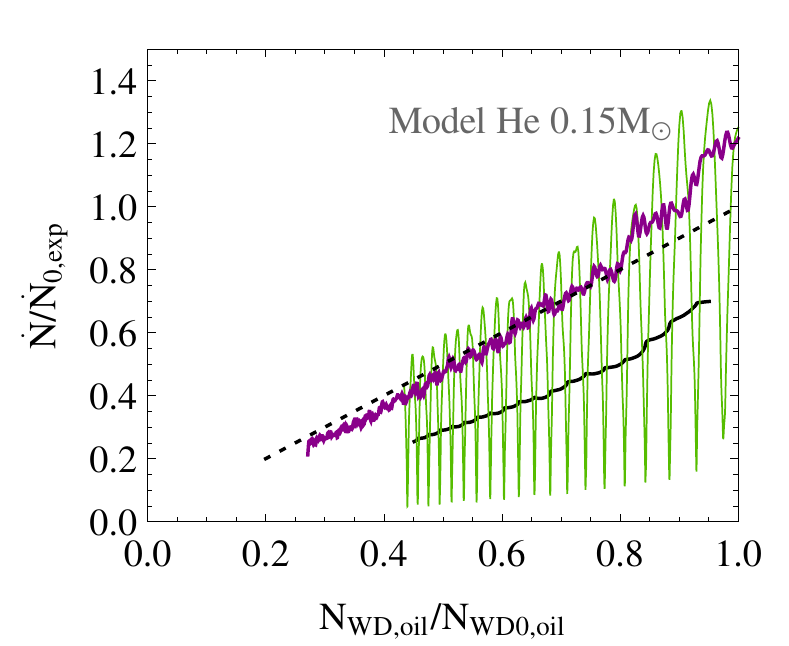}
\end{minipage}
\caption{
The rate of mass transfer, $\dot N$, plotted as a function of the number of
particles in the white dwarf's atmosphere, $N$, for two runs exhibiting strong
mass transfer. In both cases the binary contains a $0.15\,M_\odot$ helium white
dwarf in an orbit with a $1.4\,M_\odot$ neutron star. The purple line corresponds to a circular orbit at separation $1.15\,a_{\rm RLO}$. The black dashed line shows the expected behaviour if the mass transfer scaled linearly with particle number. The solid green line corresponds to a system with the same separation at the pericentre and an eccentricity $e=0.02$, and the solid black line shows a corresponding running average over one orbital period. One may see that both  circular and eccentric mass-transfer rates follow a similar trend as a function of $N$.
}
\label{fig:InstMTVersusN}
\end{figure}

The rate at which SPH particles are transferred through the $L_1$ point, $\dot{N}$, is expected to be approximately proportional to the number of Oil particles in the atmosphere, $N$, since the mass of the atmosphere is independent of the number of particles used to represent it. We tested this for a pair of long-duration runs at a fixed separation and verified that it is indeed the case, as shown in Figure~\ref{fig:InstMTVersusN}.  We also tested an eccentric system with the same pericentre separation and $e=0.02$, and find that the same trend applies. We design the main set of runs so as to avoid transferring a large fraction of the particles in order to keep the particle-transfer rate approximately constant.

\subsection{Application of model to eccentric systems}

In order to test the applicability of our model to eccentric binaries we performed a set of runs for the $0.15\,M_\odot$ He white dwarf model. The orbits were defined by the same $r_{\rm min}=1.15\,a_{\rm RLO}$, corresponding to strong mass transfer, and a range of eccentricities.  

\begin{figure*}
\includegraphics[width=2\columnwidth]{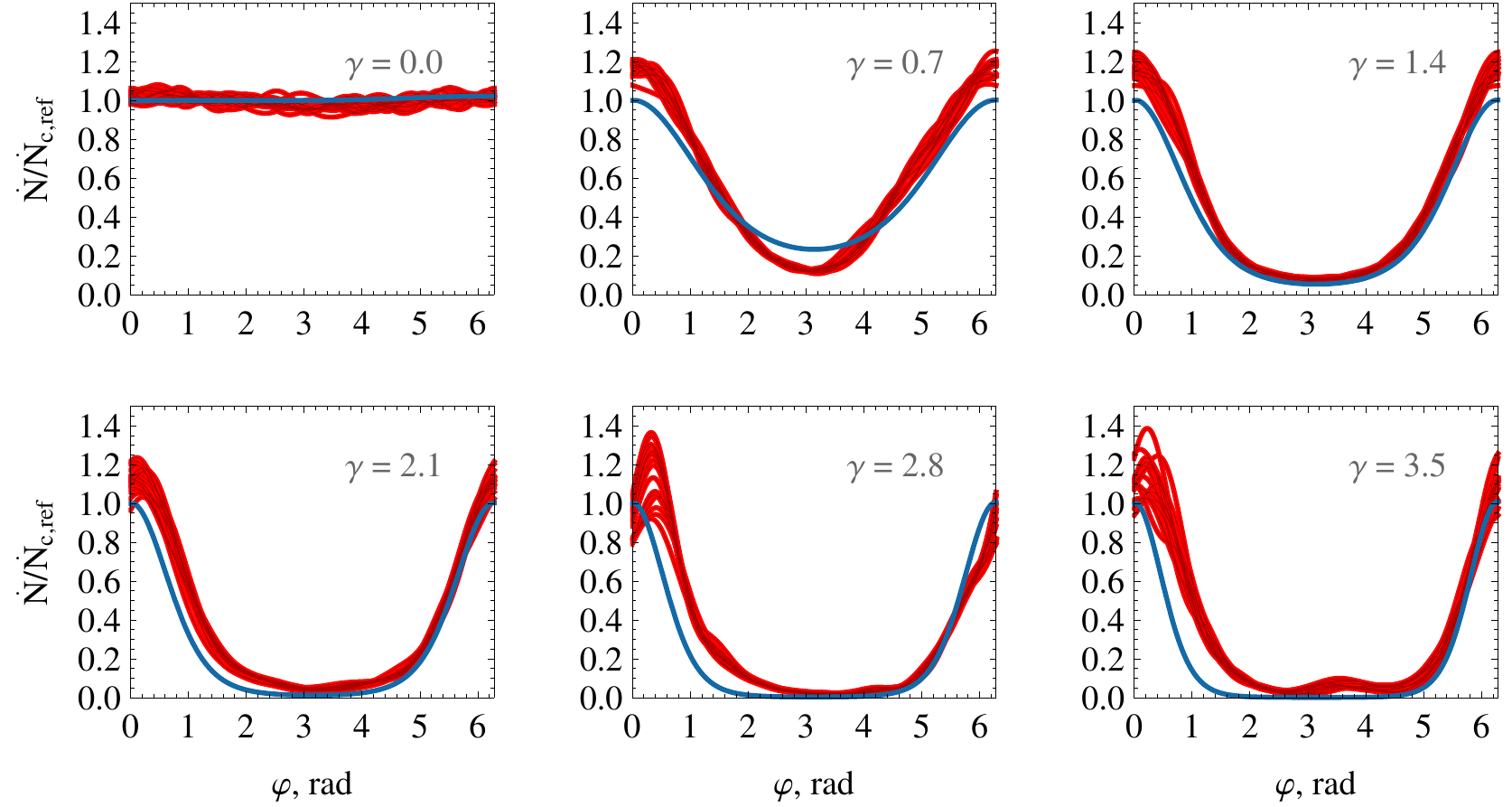}
\caption{
The plots show the ratio of the mass-transfer rate to the reference circular
value, $\dot N/\dot N_{\rm c,ref}$, as a function of orbital phase.  Different
panels show runs with different eccentricities, labelled with the parameter
$\gamma=R_{\rm WD} e/h_{\rm MT}$.
In each case the binary contains a $0.15\,M_\odot$ white dwarf and a 
$1.4\,M_\odot$ neutron star, on an orbit where $r_{\rm min}=1.15\,a_{\rm RLO}$,
corresponding to strong mass transfer at pericentre. One may see that the measured mass-transfer rate for eccentric systems (shown by the red curves) may be approximated by the circular mass-transfer rate, calculated instantaneously at the same binary separation (shown by the blue curves). We remove all the
transferred particles from the simulation once they cross the plane half way
between the $L_1$ point and the accretor to exclude back reaction effects. The width
of the top left panel stripe (circular binary) is due to small stellar proper
oscillations throughout the orbit.  These are an effect of initial conditions,
and similar oscillations can be discerned in the other panels. For higher values
of $\gamma$ we see a systematic increase in the spread of $\dot{M}$, which we interpret
as being caused by surface waves on the donor white dwarf.  
}
\label{fig:EccMTFig} 
\end{figure*}

In Figure~\ref{fig:EccMTFig} we plot the instantaneous mass-transfer rates in
our models as functions of orbital phase (red curves). 
Once normalised to remove the dependence on the number of Oil particles remaining in the envelope (see Figure~\ref{fig:InstMTVersusN}) we find that the mass-transfer rate at a given eccentricity is, to a good approximation, a function of orbital phase alone. A small spread in the curves is visible, which may be explained by tidally-powered oscillations of the white dwarf's surface, mass transfer and, in the eccentric simulations, the periodic opening and closing of the stream. To characterise these oscillations we have measured the mean radial velocity of the Oil particles on the model's surface as a function of time. In each case the frequency spectrum of the radial oscillations has a dominant mode that corresponds to the dynamical timescale of the donor, which is $0.22$ to $0.24$ times the orbital period. This leads to the small perturbations to the mass-transfer rate at a few times the orbital frequency which are visible in Figure~\ref{fig:InstMTVersusN}.  As is apparent, 
the effect of the oscillations on the mass-transfer rate, averages out over a few orbits.  
The oscillations also cause a small quantity of shock heating, of order a few times $10^{-3}$ of the atmosphere's internal energy per orbital period. The properties of surface oscillations for the other masses of the white dwarf are observed to be the same.

The uniformity of the mass-transfer rate at a given orbital phase, taken together with the observation that our circular water-only models adjust on a sub-orbital timescale, suggests that it might be possible to calculate the mass transfer rate by taking the instantaneous circular mass-transfer rates at the various separations around the orbit.  In this case the shape of the curve is parametrised by $\gamma=R_{\rm WD}e/h_{\rho}$, the depth to which the Roche lobe penetrates into the stellar surface in units of the atmospheric scale height.
We calculate this rate using the instantaneous orbital separation in Equation~\ref{MDotInst} and plot the results as the blue curves in Figure~\ref{fig:InstMTVersusN}.  It can be seen that the two agree quite well, and hence that mass-transfer rates on eccentric orbits may be calculated by using Equation~\ref{MDotInst}, as for circular systems. In other words, the donor is able to instantaneously adjust its mass-transfer rate over an eccentric orbit.

This result is important, since it tells us that we can apply our results to real WD--NS binaries.  Real white dwarfs have envelope scale heights too small for us to represent using the Oil-on-Water method.  However, these two results -- that the star re-adjusts on a shorter timescale than the orbital period, and that we can replace the mass-transfer rate in an eccentric binary with the instantaneous mass-transfer rate for a circular binary at the same separation -- allow us to calculate orbit-averaged mass transfer rates for WD--NS binaries at arbitrary separation and any small eccentricity.
We write $\dot{M}(a, e) = \dot{M}_{\rm circ}(\Delta R) f(\gamma)$, where $\Delta R$ is the separation between the white dwarf and neutron star at pericentre. For a general mass-loss law we define 
$\gamma=(R_{\rm WD} e)/h_{\rm MT}$ where  $h_{\rm MT}=\dot{M}(\Delta R)/\dot{M}^\prime(\Delta R)$ is the characteristic length-scale associated with the mass-transfer process.  Where, as here, Ritter's Law (Equation~\ref{MDotInst}) applies $h_{\rm MT} = h_\rho$ and hence $\gamma=R_{\rm WD}e/h_{\rho}$. Then, by taking an average over a Keplerian orbit with a small eccentricity $e\ll 1$, the expression for $f(\gamma)$ becomes
\begin{equation}
\label{EccEqnSPH}
f(\gamma)=\dfrac{1}{\pi} \int_{0}^{\pi} \dfrac{\dot{M}[\Delta R - \gamma h_{\rm MT}(1-\cos{\varphi})]}{\dot{M}(\Delta R)} \dd \varphi.
\end{equation}
Hence the effect of eccentricity on the mass-transfer rate may be expressed through a single dimensionless parameter $\gamma$,
which describes how deeply the Roche lobe's surface penetrates into the atmosphere on eccentric orbits in units of $h_\rho$.

In order to assess the suitability of our assumption of instantaneous mass transfer we predict the orbit-averaged mass-transfer rates $f(\gamma)=\dot{M}(\Delta R, e)/\dot{M}_{\rm orb}(\Delta R)$, and widths of the peaks of mass transfer $\Delta \varphi$ using the instantaneous assumption, and compare them to the values that we measure in the SPH simulations.  The results are presented in Table~\ref{table:EccentricMeasurements}.  We observe good agreement at low eccentricities, and moderate agreement at higher eccentricities.  The poorer match at large $e$ is most likely a result of waves excited on the surface of the white dwarf, which have amplitude approximately proportional to the eccentricity.  This effect is not expected to be relevant for real WD--NS binaries where $e\ll 1$.

\begin{table}
\caption{Measured properties of eccentric mass transfer}
\begin{minipage}{\columnwidth}
\begin{center}
\begin{tabular}{|c|c|c|c|c|c|c|c|}
\hline 
 $e$ & $\gamma$ & $\Delta \varphi_{\rm exp}$ & $\Delta \varphi_{\rm SPH}$ & $f_{\rm exp}(\gamma)$ & $f_{\rm SPH}(\gamma)$ & $f_{\rm amp}$ & $\varphi_{\rm lag}$\\ 
\hline 
$0.00$ & $0.00$ & $0.68$ & $0.68$ & $1.00$ & $1.00$ & $1.00$ & $0.00$ \\ 
$0.02$ & $0.70$ & $0.46$ & $0.41$ & $0.57$ & $0.61$ & $1.06$ & $0.01$ \\ 
$0.04$ & $1.41$ & $0.32$ & $0.32$ & $0.41$ & $0.47$ & $1.17$ & $0.05$ \\ 
$0.06$ & $2.11$ & $0.25$ & $0.28$ & $0.33$ & $0.41$ & $1.24$ & $0.13$ \\ 
$0.08$ & $2.82$ & $0.21$ & $0.24$ & $0.30$ & $0.35$ & $1.18$ & $0.35$ \\ 
$0.10$ & $3.53$ & $0.19$ & $0.24$ & $0.27$ & $0.38$ & $1.39$ & $0.19$ \\ 
\hline 
\end{tabular} 
\end{center}
Notes: Values are for a simulation with a $0.15\,M_\odot$ white dwarf, a $1.4\,M_\odot$ neutron star, and orbits with pericentre separations $r_{\rm min} = 1.15\,a_{\rm RLO}$.  The parameter $\gamma=R_{\rm WD} e/h_{\rho}$ determines how the mass-transfer rate varies with the orbital phase. $\Delta \varphi$ shows the fraction of the orbital period over which a $1\sigma$ fraction (68 per cent) of  the mass transfer takes place, centred on the $\dot{M}$ peak. $f(\gamma)$ is $\dot{M}_{\rm orb}(\Delta R, e)/\dot{M}_{\rm orb}(\Delta R)$; $f_{\rm exp}$ is the value expected from assuming an instantaneous variation of the mass-transfer rate around the orbit whereas $f_{\rm SPH}$ is the value measured from our SPH calculations. The amplification of $f$ compared to the predicted value, $f_{\rm amp}=f_{\rm SPH}/f_{\rm exp}$.  $\varphi_{\rm lag}$ is the lag in orbital phase between the $\dot{M}$ peak and the pericentre.
\end{minipage}
\label{table:EccentricMeasurements}
\end{table}

One difference between the SPH models and analytic predictions is that, in the SPH models, we observe a lag between the peak of mass transfer and the point of pericentre passage, $\varphi_{\rm lag}$, which increases with eccentricity.  The same effect was also reported by \citet{Lajoie2011}, who suggest that the lag is likely due to the dynamical free-fall timescale of the transferred material. The observed lag increases with $e$ and, extrapolated beyond $0.1$, is broadly consistent with the results of \citet{Lajoie2011}. For real WD--NS binaries, where $e\ll 1$, the lag is expected to vanish. In any case, it does not affect the orbit-averaged mass-transfer rate, which is the quantity of physical interest for the long-term evolution of the binaries.

Equation~\ref{EccEqnSPH} permits several simple expressions for orbit-averaged mass-transfer rates in eccentric systems. For an ideal gas atmosphere, the case that we present in Figure~\ref{fig:EccMTFig}, $f(\gamma)=I_0(\gamma)e^{-\gamma}$, where $I_0$ is the modified Bessel function. For systems that are sufficiently eccentric to have large $\gamma$ but still also have small $e$ the dependency is simply $f(\gamma)=1/\sqrt{2\pi\gamma}$ for any mass loss law. In this case a $1\sigma$ fraction of the material is transferred within the angle $\pm\Delta \varphi = \pm 1/\sqrt{\gamma}$ of the orbit. The result suggests that even for sharp mass-transfer profiles the eccentricity $e$ still enters the expression for the orbital averaged mass-transfer rates, at odds with the assumption made by \citet{Sepinsky2007} and \citet{Davis2013a}.

Finally, we note that in real WD--NS systems mass loss through the $L_2$ point is not expected. This is because the minimum distance between the $L_1$ and $L_2$ equipotential surfaces is about $0.1\,R_{\rm WD}$, which is much greater than the atmospheric scale height, $h_{\rho}$. In our models, however, we observe marginal mass loss through $L_2$ for the $0.15\,M_\odot$ white dwarf, but only because the Oil atmospheres have artificially high $h_{\rho}$.

\subsection{Connection of simulations to WD--NS systems}

\label{sect:connection}

Armed with the conclusions from the previous section, we construct a set of runs that we use to model mass transfer in real WD--NS binaries.   
The Oil-on-Water method requires us to use artificially exaggerated atmospheres,
with $h_{\rho}$ being a few per cent of the white dwarf's radius (Section~\ref{sec:SingleSetup}). The
characteristic scale heights $h_{\rho}$ of real white dwarfs are much smaller,
typically between about $10^{-3}\,R_{\rm WD}$ and $10^{-5.5}\,R_{\rm WD}$. 
However, since we have shown that the
mass-transfer rate (Equation~\ref{MDotInst}) depends only on the number of scale heights by which
the Roche lobe digs into the white dwarf's atmosphere, rather than just on
$h_{\rho}$, by simulating a binary with $h_{\rho}$ of $0.03\,R_{\rm
WD}$ and $e=0.03$ we are able to model, for example, a real O-Ne white dwarf
with $h_{\rho} = 10^{-5}\,R_{\rm WD}$ and $e=10^{-5}$.
In the first case the binary separation needs to vary appreciably over the orbit to alter the mass-transfer rate. In the second case a small eccentricity is enough to affect the mass-transfer rate, since the atmospheric scale height of the donor is much smaller.

Our model allows us to model systems with eccentricities between zero and a few times the critical eccentricity $e_{\rm crit}$ for mass transfer to turn on and off during the orbit. Given the scales of our model atmospheres the relationship between the physical eccentricities $e_{\rm phys}$ and corresponding model eccentricity, 
$e_{\rm model} = 0.03\,e_{\rm phys}/e_{\rm crit}$, where $e_{\rm crit}$ can be calculated from Equation~\ref{eqn:ecrit}.  The critical eccentricities corresponding to the white dwarfs that we model are given in Table~\ref{tab:eCrit}, where we assume typical tidally-heated atmosphere temperatures of $4\times 10^4\,{\rm K}$.

\begin{table}
\caption{Critical eccentricities for our model white dwarfs}
\begin{tabular}{lll}
\hline
Mass \hspace{1.5cm} & Composition \hspace{1.cm} & $\log_{10}e_{\rm crit}$ \hspace{0.5cm} \\
${\rm M}_\odot$ \\
\hline
0.15 & He & -3.7 \\
0.6  & CO  & -4.7 \\
1.0  & ONe & -5.1 \\
1.3  & ONe & -5.5 \\
\hline
\end{tabular}
\label{tab:eCrit}
\end{table}

\begin{figure}
\includegraphics[width=\columnwidth]{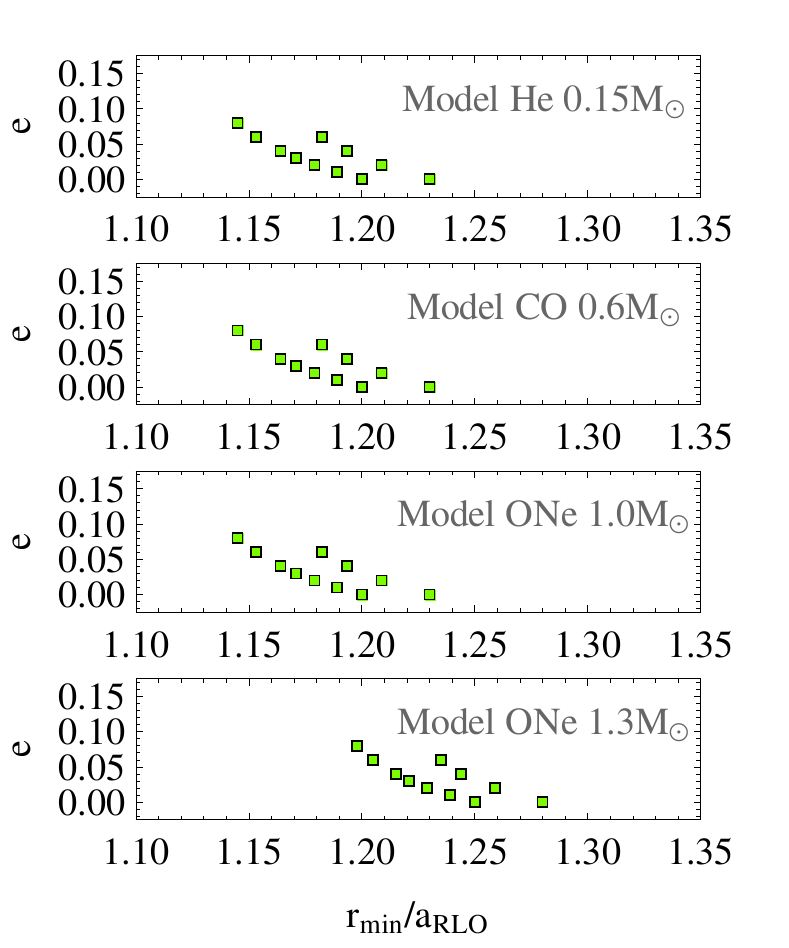}
\caption{Our main set of $44$ eccentric simulations in the $(r_{\rm min},e)$ plane. The orbital separation at the pericentre, $r_{\rm min}$, is expressed in units of $a_{\rm RLO}$, which corresponds to the separation for Roche lobe overflow in circular systems. We arrange the initial conditions to obtain approximately the same $\dot{M}$ rate for each stripe. The simulations ran over at least $20$ binary orbits, entering a quasi-stationary state after $5-10$ orbits. The initial conditions for the $1.3\,M_\odot$ system are offset from the others to compensate for strong donor expansion, discussed in Section~\ref{sec:BinSetup}.}
\label{fig:MainRuns}
\end{figure}

Our main set of models comprises $44$ runs, whose orbital elements are presented in Figure~\ref{fig:MainRuns}.  The longest stripes in the figure correspond to the main runs, which we run for 20-30 orbital periods each. The systems transfer approximately one third of their atmospheres over $20$ orbits. The runs represented by shorter stripes to the right in the figure correspond to lower mass-transfer rates, which we use to verify that our results still apply at these rates.   

\section{SPH simulations: results}

\label{sec:Flows}

In this Section we discuss the general properties of the accretion flows that we see in our models and use the results to extract physical quantities to use in a long-term evolution model.

\begin{figure*}

\begin{minipage}{0.98\columnwidth}
\includegraphics[width=\columnwidth]{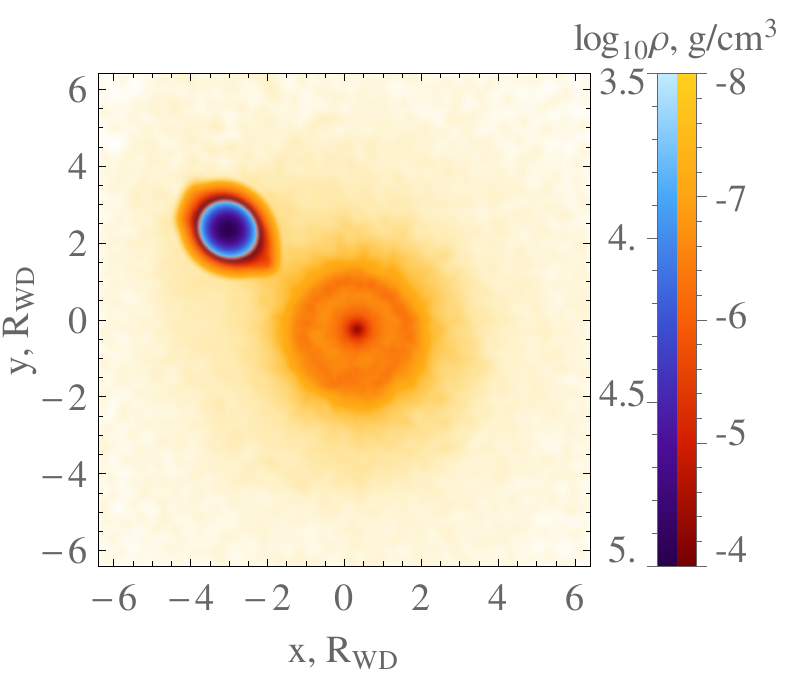}
\end{minipage}
\begin{minipage}{0.98\columnwidth}
\includegraphics[width=\columnwidth]{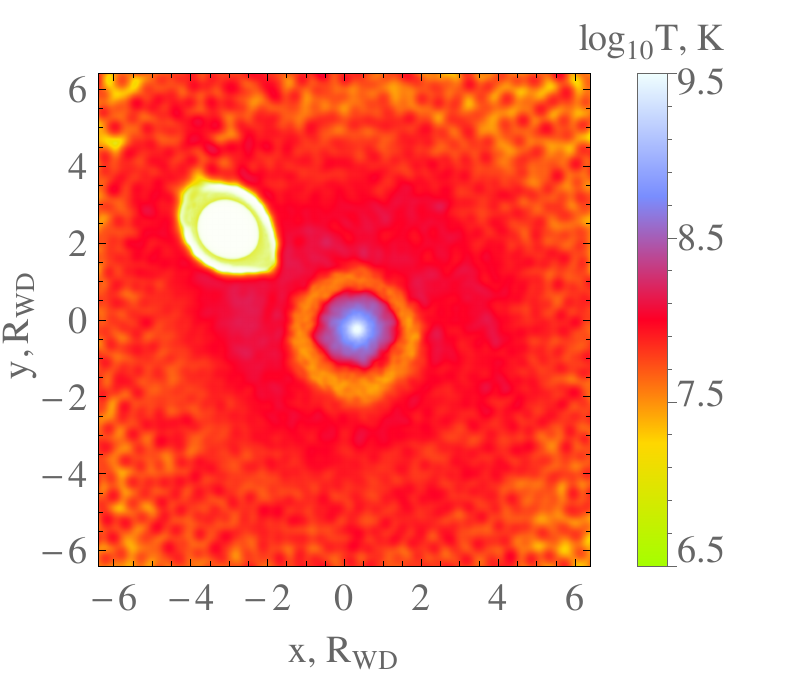}
\end{minipage}

\begin{minipage}{0.98\columnwidth}
\includegraphics[width=\columnwidth]{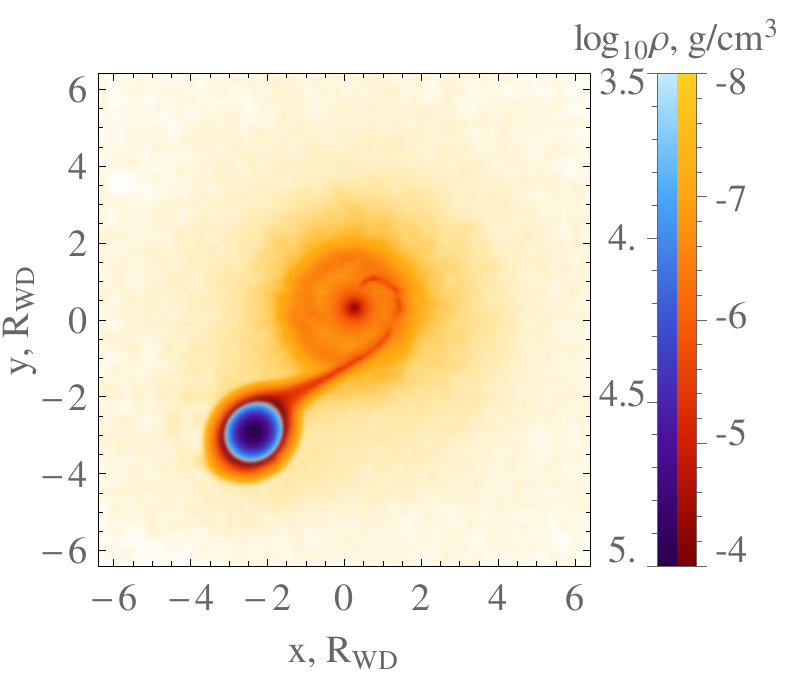}
\end{minipage}
\begin{minipage}{0.98\columnwidth}
\includegraphics[width=\columnwidth]{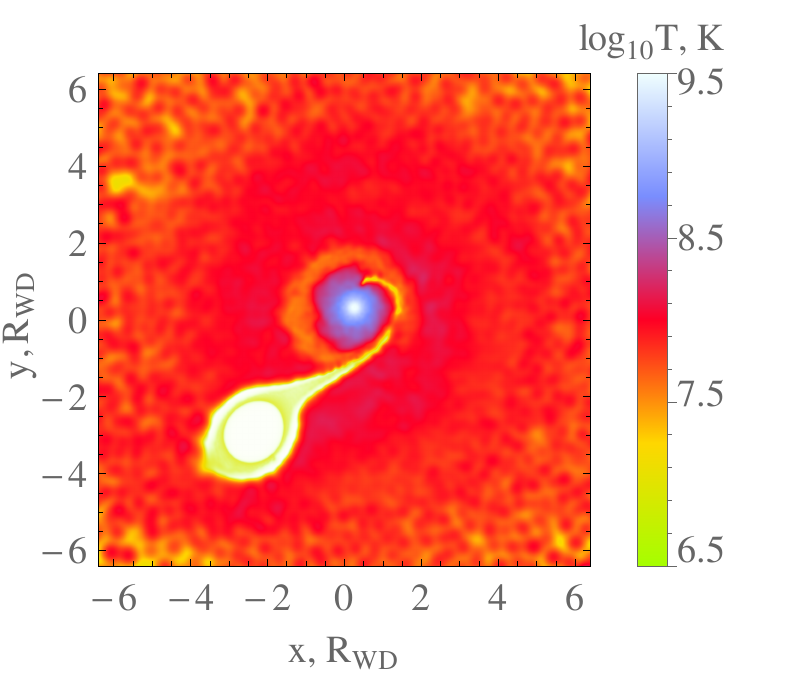}
\end{minipage}

\begin{minipage}{0.98\columnwidth}
\includegraphics[width=\columnwidth]{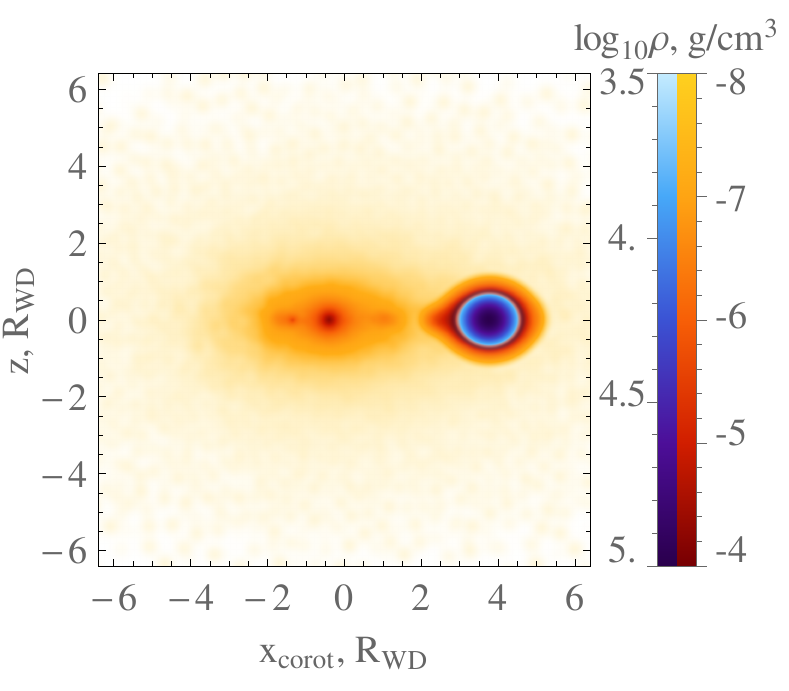}
\end{minipage}
\begin{minipage}{0.98\columnwidth}
\includegraphics[width=\columnwidth]{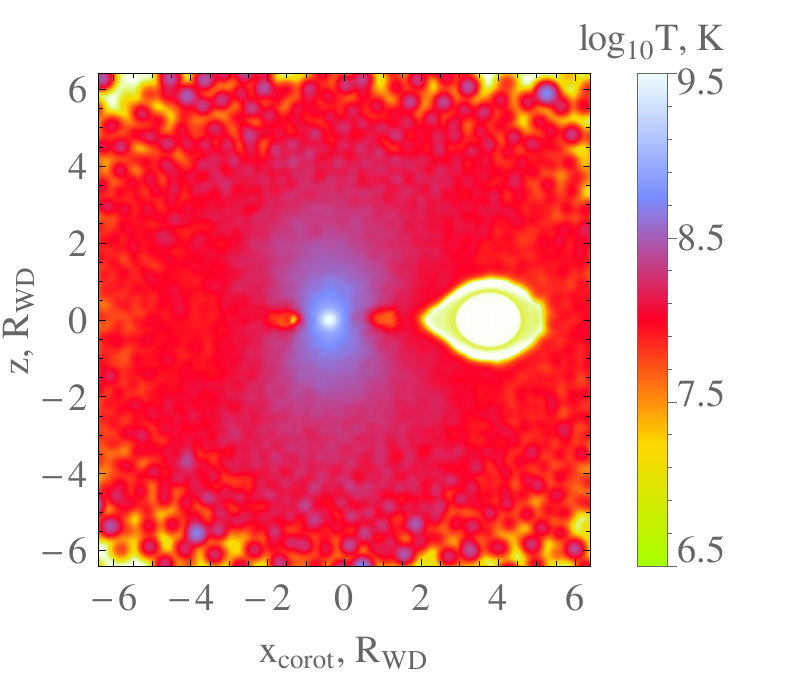}
\end{minipage}

\caption{Snapshots of mass flows in a binary with a $0.15\,M_\odot$ WD, a $1.4\,M_\odot$ NS, and $e=0.04$. 
The top two panels show density (left)
and temperature (right) sections in the orbital plane.  The bottom row shows
density (left) and temperature (right) in a plane that is perpendicular to the orbital
plane and goes through the centres of the two stars.  
The snapshots are shown after 13 orbits, $1/8$ of an orbit before (top panels) and after (middle
and bottom panels) pericentre passage. 
At this stage the disc has formed and entered a steady state.
The density plots show eccentric structures in the disc, the complex character of
the flow near the circularisation radius and a strong density cusp near the NS. 
The envelope surrounding the binary is sparse but its total mass is significant compared to the disc.
The temperature plots show that our artificially heated WD atmosphere is
cold compared to both the disc and the envelope.
}
\label{fig:MTMaps}
\end{figure*}

Example snapshots of the density and temperature structure of one of our simulations can be seen in Figure~\ref{fig:MTMaps}. The features seen in that figure give a good qualitative picture of the mass flows that our models exhibit. The systems in our simulations quickly become non-conservative and experience strong mass loss.

\begin{figure}
\includegraphics[width=\columnwidth]{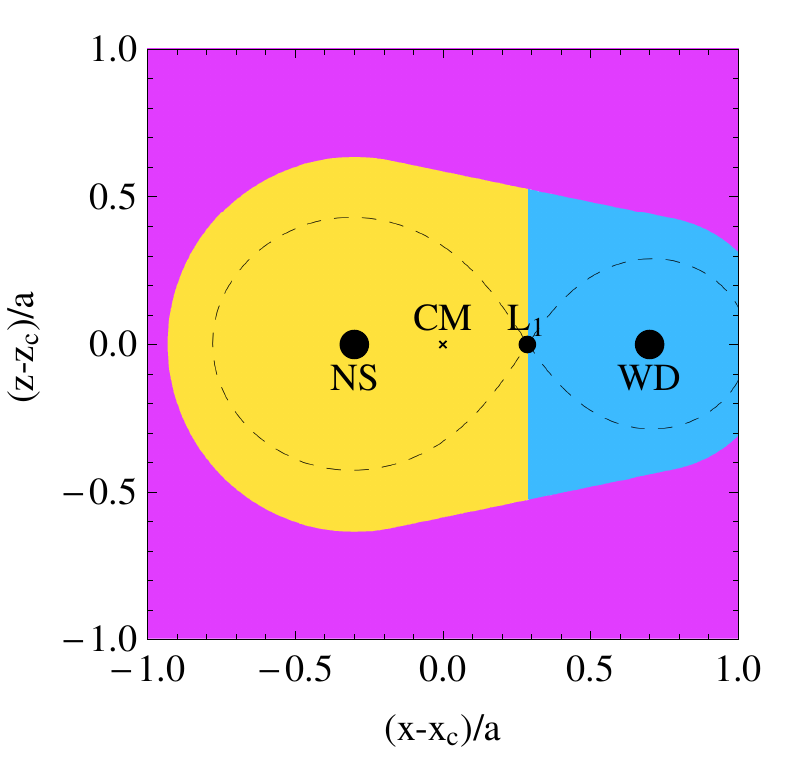}
\caption{We classify SPH particles according to the region of space they are in. The WD region (cyan) is particles associated with the white dwarf, the NS region (yellow) is particles associated with the neutron star, and the CE region (magenta) is particles associated with the common envelope, as opposed to either star in particular. The figure shows a section in a plane through the centres of the two stars and perpendicular to the orbital plane.
The division is done by constructing a convex hull tiled over two spheres centred on the white dwarf and the neutron star, each sphere having a radius of $1.4$ times the star's Roche lobe
radius. The WD region is separated from the NS region by a plane passing through the $L_1$ point perpendicular to the binary axis.
}
\label{fig:MTScheme}
\end{figure}

In order to probe the accretion flows in our simulations it is necessary to identify whether SPH particles are still attached to the white dwarf, have been transferred to the neutron star, or have been lost from the two bodies into a common envelope surrounding the binary.  We divide the particles up using the scheme illustrated in Figure~\ref{fig:MTScheme}. The geometric (computationally inexpensive) division is done by using a convex hull tiled around two spheres centred on the white dwarf and the neutron star. The spheres are set to have empirically chosen radii of $R_{\rm div} = 1.4\,R_{\rm RL}$ and the Roche lobe radii are calculated instantaneously. The resulting ovoid is split in two by a plane perpendicular to the binary axis and passing through the $L_1$ point. This allows us to measure the properties of material in the different regions, and the flows between them.

\begin{figure}
\includegraphics[width=\columnwidth]{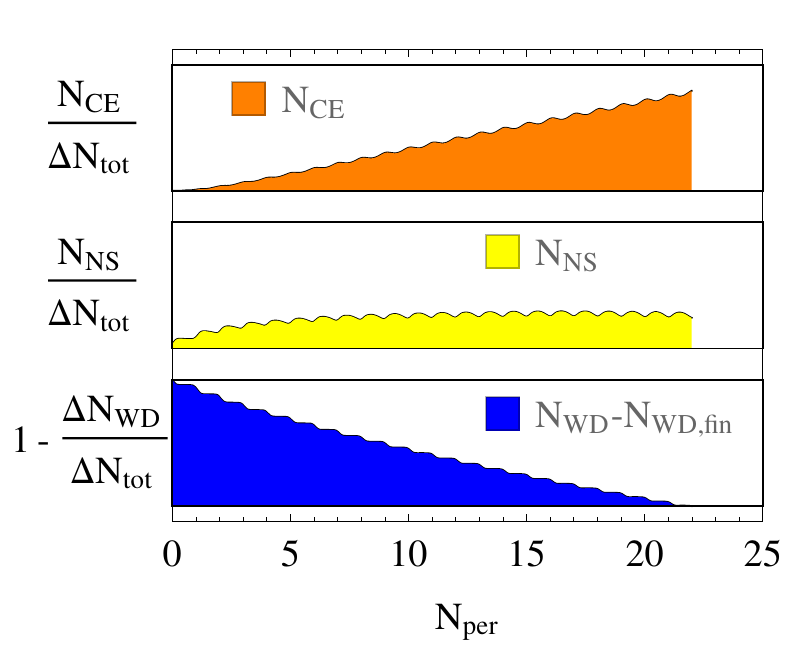}
\caption{
Evolution of the number of particles in each of the three zones in units of $\Delta N_{\rm tot}$, the total number of particles transferred during the simulation as a function of time in units of the orbital period, for an example system with a $0.15\,M_\odot$ helium white dwarf donor on
an eccentric orbit with $e = 0.08$. The three panels show the number of
particles in the common envelope zone (top), the number bound to the neutron
star (middle) and the number bound to the white dwarf (bottom). 
Particles that remain bound to the white dwarf at the end of the run are
subtracted off in the third figure. One can see that the mass transfer takes
place in short bursts around pericentre. The disc growth saturates after about
$10$ orbits; thereafter the transferred mass is lost to the common envelope.
The plots are normalized by the total number of transferred particles, which in
this case is $1.2\times 10^5$, amounting to about one third of all the Oil
particles in the white dwarf's atmosphere.
}
\label{fig:MTProgress}
\end{figure}

The evolution of the masses in the different zones as functions of time for a typical model is plotted in Figure~\ref{fig:MTProgress}.
The rate of flow of mass in to the disc through the L1 point is roughly constant throughout the simulation when averaged over a whole orbital period. After about ten orbital periods, mass is lost in a wind from the disc at a constant rate that is approximately equal to the rate at which it is gained from the donor. The disc mass thereafter remains nearly constant and the flow becomes quasi-steady.  We ensure that the systems have entered this steady regime before extracting orbit-averaged quantities.

We have tested the effect of our choice of radii for the spheres in the identification scheme above.  If we use a smaller radius than $1.3\,R_{\rm RL}$, parts of the atmosphere and disc are misidentified as lost particles during pericentre passages in the most eccentric runs. Using larger radii does not have an appreciable effect on the identification, except for reducing the number of particles identified as lost. An alternative approach would be to use a criterion based on binding energy, however we find that this is less suitable for our models.  Particles located outside the instantaneous Roche lobe, for example, would be identified as unbound, whereas in practice they may spend appreciable amounts of time on the surface of the white dwarf before being transferred to the neutron star. Similarly, a fraction of the particles in the disc would be misidentified as lost due to them acquiring a high kinetic energy well before they are actually ejected.

\subsection{Disc properties}

\begin{table}
\begin{minipage}{\columnwidth}
\caption{Summary of measured disc properties}
\begin{center}

\begin{tabular}{l c c c c}
\hline
Property & $0.15\,M_\odot$ &  $0.6\,M_\odot$ &  $1.\,M_\odot$ &  $1.3\,M_\odot$\\
\hline
(Spatial res)/$100\,\textrm{km}$ & $3.70$ & $2.07$ & $1.14$ & $0.67$\\
$\sigma_R/r_{\rm circ}$ & $1.68$ & $2.15$ & $2.16$ & $2.28$\\
$\sigma_z/\sigma_R$ & $0.51$ & $0.50$ & $0.50$ & $0.49$\\
$\langle t_{\rm NS}\rangle/P$&$5.62^{\pm 0.27}$&$4.56^{\pm 0.27}$&$4.72^{\pm 0.36}$&$4.63^{\pm 0.50}$\\
$\sigma(t_{\rm NS})/P$ &$2.91^{\pm 0.17}$&$2.99^{\pm 0.25}$&$3.28^{\pm 0.35}$&$3.28^{\pm 0.49}$\\
$\tau_{\dot{M}}/P$&$7.6$&$6.3$&$6.8$&$6.9$\\
$\tau_{\rm visc}/P$&$22.8$&$14.2$&$11.0$&$11.0$\\
$\tau_{\rm th,inner}/P$&$6.7$&$6.8$&$7.0$&$6.9$\\
\hline
\end{tabular}
\end{center}
Notes: The spatial resolution is the best value in the set of runs
for a given model. The RMS radial disc size $\sigma_R$ is measured relative to the
circularisation radius $r_{\rm circ}$ and is measured at the end of the simulation, as is the RMS vertical extent $\sigma_z$.  Timescales are given
relative to the orbital period $P$.  $\langle t_{\rm NS}\rangle$ is the mean time
material spends in the neutron star zone before being ejected, whilst $\sigma(t_{\rm NS})$ is the $1\sigma$ spread in $t_{\rm NS}$ for individual particles.  The residuals in the quantities represent how they vary between different runs in Figure~\ref{fig:MainRuns}. The disc replenishment time, $\tau_{\dot{M}} = M_{\rm disc}/\dot M$.  The characteristic viscous timescale for an alpha-disc, $\tau_{\rm visc}$, is calculated given the measured disc thicknesses and $\alpha=0.01$.  The heating timescale
$\tau_{\rm th,inner}=T/\dot{T}$ for material locked to the neutron star.
\label{table:SPHResultsDisc}
\end{minipage}
\end{table}

The disc in our SPH models is thick and corresponds to the super-Eddington accretion
regime.  This is in part because we do not remove SPH particles from near the
neutron star. Since we do not include radiative cooling, the modelled accretion flow is radiatively inefficient. This is physically realistic for super-Eddington mass-transfer rates \citep{Scadowski2014}, however, because we do not include radiative cooling the temperatures seen in the runs are probably exaggerated. Since the simulation physics does not allow for jet formation, mass is lost to the common envelope purely through hot disc winds, as seen in Figure~\ref{fig:MTMaps}.

The disc is observed to enter a steady state after about $10$ orbital periods. A fraction of the inner disc remains locked to the neutron star, being replenished slowly and getting gradually heated. The cold stream reaches down to the circularisation radius, where it fragments into a persistent eccentric cool ring. Since the temperature of the freshly transferred material significantly exceeds the temperature of the white dwarf's atmosphere, the latter is effectively cold compared to the disc. The outflows from the disc are seen to be generally much hotter than the stream. Most of the disc mass is located within $\sim 2\,r_{\rm circ}$. Simulations with the same orbit-averaged mass-transfer rate but different orbital eccentricities have very similar discs.

We measure several characteristic timescales for the disc, each being of order
$10\,P$; the values are given in Table~\ref{table:SPHResultsDisc} along with some other properties of the disc. One 
timescale comes from the mean time SPH particles spend in the neutron star zone
before being lost to the common envelope.  Another timescale corresponds to the
time it takes for the disc to be completely replenished.  
Another very similar timescale is the characteristic heating timescale $T/\dot T$.
Finally, we obtained viscous timescales for the disc from the $\alpha$-disc
model, assuming $\alpha_{\rm disc}=0.01$ and estimating $H/R$ as the ratio of
standard deviations in $z$ and radial positions $r$ with respect to the neutron
star.

\begin{table}
\begin{minipage}{\columnwidth}
\caption{Summary of measured mean properties of the ejected material.}
\begin{center}
\begin{tabular}{l c c c c}
\hline
Property & $0.15\,M_\odot$ &  $0.6\,M_\odot$ &  $1.0\,M_\odot$ &  $1.3\,M_\odot$\\
\hline
$T_{\rm ej}/(10^8\,\textrm{K})$&$1.16^{\pm 0.35}$&$3.90^{\pm 0.89}$&$6.46^{\pm 1.49}$&$11.60^{\pm 2.92}$\\
$\epsilon_{\rm ej}/(E_{12}/\mu)$&$1.88$&$1.91$&$1.78$&$1.67$\\
$j_{\rm ej}/(J_{12}/\mu)$&$0.50$&$0.64$&$0.76$&$0.84$\\
\hline
\end{tabular}
\end{center}
Notes: All the quantities are measured for the material at the moment of entering the common envelope zone. These quantities represent the averages over the particles that escape, after the systems have entered a steady regime. $T_{\rm ej}$ is the temperature of ejected material. The specific angular momentum $j_{\rm ej}$ and the binding energy $\epsilon_{\rm
ej}$ of the ejected material are measured with respect to the orbital quantities per
unit reduced mass  $\mu$.
\label{table:SPHResultsCE}
\end{minipage}
\end{table}

\subsection{Parameters for modelling the long-term binary evolution}

We measure the mean specific binding energy, angular momentum and temperature of the material as it is lost to the common envelope zone for the last time. The numerical values are given in Table~\ref{table:SPHResultsCE}. We found very little dependence of the parameters on time in any of the runs, and almost no dependence on $\dot{M}$ or $e$. This is consistent with the disc properties measured earlier. Since the material spends on average about five
binary orbital periods in the disc before being ejected, the  details of how it was transferred are expected to have little connection to its properties when it is lost. The values that we measure are therefore likely to be similar in systems with any eccentricities much less than unity, which includes all the real WD--NS systems when they come into contact.  Binaries with eccentricity comparable to unity could be dynamically different from the systems in the present study, which would probably have a bearing on the properties of the lost material \citep{Regos2005, Lajoie2011}.

We measure the gravitational binding energy of the lost material, which we use in Section~\ref{sec:DiscussionBinaryEvolModel} to
discuss the common envelope energetics. The specific angular momentum of the
lost material, $j_z$, is dynamically important for the long-term evolution of
the binary, as we discuss in Section~\ref{section:stab}.

\label{EccMTGen}

The long-term evolution of WD--NS binaries is sensitive to the amounts of mass and angular momentum lost during mass transfer.  We follow \citep{1982ApJ...254..616R} and define two parameters to describe how non-conservative the evolution is, which we measure directly from our simulations. The parameter $\beta$ is defined as the fraction of the transferred material which gets accreted:
\begin{equation}
\beta=-\dot{M}_{NS}/\dot{M}_{WD}
\end{equation}
Since we do not allow for accretion of gas by the neutron star, the actual values of beta in our simulations correspond to the fraction of material that accumulates in the disc and hence approach zero.   Since our models represent the highly super-Eddington phase of evolution, if the accretion rate on to the neutron star is limited to the Eddington rate then this is realistic for real systems.

The second parameter, $\alpha$, is defined as the specific angular momentum carried away by the lost material, 
\begin{equation}
\label{AlphaDef}
\alpha\equiv \dfrac{\dot{J}_{\rm loss}}{\dot{M}_{\rm loss}}/\dfrac{J_{12}}{\mu},
\end{equation}
where the reduced
mass of the binary $\mu=(M_{\rm WD} M_{\rm NS})/(M_{\rm WD} + M_{\rm NS})$.  
In the simulations we find that the specific angular momentum of material lost from the disc reaches a constant value after about ten orbital periods which does not change subsequently, independently of the length of the run. Hence we take $\alpha$ equal to the normalised values of specific angular momentum of material lost from the disc provided in Table~\ref{table:SPHResultsCE}. We find that the values of $\alpha$ are well fit as a linear function of the mass ratio:
\begin{equation}
\alpha = 0.46 + 0.42 \frac{M_{\rm WD}}{M_{\rm NS}}.
\label{AlphaFit}
\end{equation}
In order to test that this result is resolution-independent we verified that the amount of specific angular momentum measured at lower mass transfer rates, in the runs represented by shorter stripes in Figure~\ref{fig:MainRuns}, agrees closely with the values measured in the regular runs. These  values are similar to those expected from a simple physical picture, as discussed further in discussion Section~\ref{section:stab}.

\section{Discussion}
\label{sec:DiscussionMain}

Our SPH simulations allow us to measure the specific angular momentum of material lost by a WD--NS binary through disc winds. We use these measurements to construct a model of the long-term evolution of these binaries,
as explained in this section. We are thus able to model the long-term evolution of a set of WD--NS binaries (i.e.
for white dwarfs of various masses). As will be shown below, we find that the mass-transfer is stable
for sufficiently low-mass white dwarfs. Our model and its parameters are described below; subsequently 
we discuss the
implications of results from our SPH runs for the model and the implications of our model for observations.

The idea of using the results from  hydrodynamic simulations to model the long-term evolution of mass transfer was proposed by \citet{Lajoie2011}. Together with \citet{Church2009} they were the first to simulate non-degenerate eccentric binaries with realistically low mass-transfer rates. The long-term evolution of stably transferring WD--NS binaries based on an empirical prescription for angular momentum loss was originally studied by \citet{Savonije1986} and recently in detail by  \citet{2012A&A...537A.104V}. The long-term evolution of both stable and unstable double WD systems was explored by \citet{Marsh2004} and \citet{Gokhale2007}. The long-term evolution of strongly eccentric general binaries has been studied by  \citet{Sepinsky2009} and \citet{Dosopoulou2016a, Dosopoulou2016}, where the authors explored the long-term effects of eccentricity. In the following we make use of the measured values for the angular momentum loss efficiency in eccentric WD-NS binaries to construct the long-term models of their evolution in both stable and unstable scenarios.

Our hydrodynamic modelling shows that circular and eccentric systems likely follow the same long-term evolution. The amount of specific angular momentum lost in the disc winds does not depend on eccentricity, but only on the mass ratio of the binary. This is likely related to the viscous timescale of the disc being long enough to erase the information about the details of how material is injected into the disc.  It should hold for all realistic WD--NS binaries at contact, where $e\ll 1$. Even though the simulations were carried out only in about $1\,\textrm{dex}$ window of eccentricities, we argue in Appendix~\ref{MDotMotivation} that this should hold true for any eccentricities so long as they are much smaller than unity. 

The eccentric contribution may be factored out from the mass-loss law (Equation~\ref{EccEqnSPH}). Since the eccentric part enters the mass-loss law as a pre-factor, it does not affect the long-term evolution \citep{Webbink1985} unless the eccentricity varies on a timescale which is comparable to or shorter than the timescale of the evolution of the mass-transfer rate. At very small eccentricities a very small fractional change in the orbital separation $a$ has the same effect as a large fractional change in $e$.  Therefore, we expect changes in $a$ to dominate the long-term evolution, which allows us to neglect the eccentricity evolution.

Hence we focus on circular systems here as they are very likely to have similar evolutionary tracks to 
the real systems possessing (very small) eccentricities.

\subsection{Binary evolution model}
\label{sec:DiscussionBinaryEvolModel}
We start with a binary shortly before it comes into contact. At this stage the mass-transfer rate is sufficiently small to be dynamically insignificant and the orbital evolution is dominated by gravitational wave emission. The rate at which angular momentum is radiated away in gravitational waves is given by \citet{Peters1964} as:
\begin{equation}
\dot{J}_{\rm GR} =
-\frac{32}{5}\dfrac{G^{7/2}M_{\rm WD}^2 M_{\rm NS}^2 (M_{\rm WD}+M_{\rm NS})^{1/2}}{c^5 a^{7/2}}.
\end{equation}
As the binary spirals together the white dwarf starts to transfer mass to the neutron star.  The mass-transfer rate is given at all times by Ritter's formula (Equation~\ref{MDotInst}), where we take the outer layers of the white dwarf to have an ideal-gas atmosphere. The mass-radius relation for the white dwarf is based on a dense set of theoretical stellar profiles obtained as described in Section~\ref{sec:SingleSetup}. What happens to the mass lost by the white dwarf depends on the mass-transfer rate, which we consider in terms of the limiting Eddington rate,
\begin{equation}
\label{eq:MDotEdd}
\dot{M}_{\rm Edd} = \frac{4\uppi G\mu_{\rm e} M_{\rm NS}m_{\rm p}}{\sigma_{\rm T}\eta_{\rm
acc}c}
\end{equation}
where $m_{\rm p}$ is the proton mass, $\sigma_{\rm T}$ is the Thomson cross-section,
the mass per electron in atomic mass units $\mu_{\rm e}\simeq 2$, and we take the accretion efficiency $\eta_{\rm acc}=0.1$.

\subsubsection{Phase 1: Sub-Eddington mass transfer}

For $\left|\dot M_{\rm WD}\right|<\dot{M}_{\rm Edd}$, we assume conservative mass transfer; that is, all the mass lost by the white dwarf is accreted by the neutron star, with the angular momentum being retained in the orbit due to viscous torques from the accretion disc.

\subsubsection{Phase 2: Jet-like outflow}

When mass transfer is only slightly super-Eddington, we assume that the
excess mass will be lost from an energetic outflow from very close to the
neutron star; i.e. a jet \citep{Fender2012}. This is analogous to the hard state of an {\it X}-ray
binary.  We treat the binary as being in this phase when $1 < \left|\dot M_{\rm
WD}/\dot{M}_{\rm Edd}\right|<(1+f_{\rm jet})$. We take the arbitrary parameter $f_{\rm jet}=1$, as motivated in the following section. During this phase the excess mass is lost
from the system with specific angular momentum equal to the specific orbital
angular momentum of the accreting neutron star \citep{Tauris1999}.

\subsubsection{Phase 3: Momentum-driven outflow}

\label{sec:DiscPhase3}

At higher mass-transfer rates we assume that, instead of forming a thin disc that accretes efficiently as in phases 1 and 2, the majority of the excess material will be lost from the disc in a slow, mechanical wind.  This will lead to a build up of a cloud around the binary, which we refer to as the common envelope since it is not bound specifically to either star. This is  what we see in the SPH simulations.

Mechanical winds are observed in real systems, e.g. high-mass {\it X}-ray binaries such as SS~433 \citep{Begelman2006}. They are expected theoretically because the disc structure should change when the accretion rates sufficiently exceed the Eddington rates \citep{Abramowicz2013}. This is also observed in simulations \citep{Scadowski2014}. The mass-transfer rate corresponding to the beginning of Phase~3 is determined by $f_{\rm jet}$ and is expected to exceed the Eddington rate by a factor of a few. The exact value of $f_{\rm jet}$ has little effect on stability of mass transfer, since most WD--NS binaries reach rates which exceed the Eddington limit by orders of magnitude \citep{2012A&A...537A.104V}. We discuss this further in Section~\ref{section:stab}.

We test whether the jetted outflow is strong enough to penetrate the cloud material. This material has an average escape velocity of:
\begin{equation}
v_{\rm esc,CE}=\sqrt{\frac{2G(M_{\rm WD}+M_{\rm NS})}{R_{\rm CE}}},
\label{eqn:vEscCe}
\end{equation}
where $R_{\rm CE}$ is the effective radius of the common envelope.  Reasonable values for $R_{\rm CE}$ are of order $a$, since the common envelope
is about the same size as the binary.  By this definition $R_{\rm
CE}=2a/\bar{\epsilon}_{\rm ej}$, where $\bar{\epsilon}_{\rm ej}=\epsilon_{\rm ej}/(E_{12}/\mu)$ and is given in Table~\ref{table:SPHResultsCE}. One can see from the table that the values of $\bar{\epsilon}_{\rm ej}\approx 1.8$ with no clear trend with
$M_{\rm WD}$: hence we take $R_{\rm CE}=1.1\,a$ in our standard run.

We assume that the jet is incident on a fraction $f_{\rm funnel}$ of the
surrounding envelope, which causes a funnel-shaped outflow as long as it has
sufficient momentum to penetrate the envelope.  The rate at which mass flows
into the funnel 
$\dot M_{\rm funnel} = f_{\rm funnel}(\dot M_{\rm disc} + M_{\rm CE}/\tau_{\rm
CE}$).  The first term in parentheses,
$\dot M_{\rm disc}=|\dot M_{\rm WD}|-\dot M_{\rm
NS}-\dot M_{\rm jet}$ is the rate at which mass is lost from the disc by the
mechanical wind, and $M_{\rm CE}$ is the mass which has built up in the common
envelope.  The second term arises because the material in the common envelope
will redistribute itself into the funnel on the dynamical timescale of the common
envelope $\tau_{\rm CE}$.  Finally, the criterion for the momentum of the jet to
penetrate the envelope becomes $\dot M_{\rm
jet}v_{\rm jet} > (\dot M_{\rm jet} + \dot M_{\rm funnel})v_{\rm esc, CE}$.
Here the mass-loss rate in the jet $\dot M_{\rm jet}=f_{\rm jet}\dot M_{\rm
Edd}$, and, since it is launched from close to the neutron star we take $v_{\rm
jet}=v_{\rm esc}(R_{\rm jet})$, where we choose $R_{\rm jet}=10^{-4}\,R_\odot$.

The material lost from the binary in the jet carries away just the specific angular
momentum of the accreting neutron star as before, but the material lost in the
disc wind additionally carries angular momentum from the disc.  We write the momentum carried off in terms of the mass ratio $q=M_{\rm WD}/M_{\rm NS}$ and the specific
orbital angular momentum, $j_{\rm orb}=J_{12}/\mu$.
For our default model we use the specific 
angular momentum in disc winds measured in our simulations. Then
\begin{eqnarray}
\dfrac{\dot J_{\dot M}}{j_{\rm orb}} & = & -\left(\dfrac{q}{1+q}\right)^2\dot M_{\rm jet} -\alpha(q)\,\dot M_{\rm disc},
\label{eqn:jdot1}
\end{eqnarray}
where $\alpha(q)$ is given by Equation~\ref{AlphaFit}.

As an alternative, we test the effect of assuming that 
the disc wind carries away the specific angular momentum of material in a Keplerian orbit at the circularisation radius.  
Following \citet*{Frank2002}, we obtain the total rate at which angular momentum is carried off by the mass lost as
\begin{eqnarray} \nonumber
\dfrac{\dot J_{\dot M}}{j_{\rm orb}} & = & -\left(\dfrac{q}{1+q}\right)^2\dot M_{\rm jet} - \left[\left(\frac{q}{1+q}\right)^2 + \right.  \\
                && \left.  +(0.500-0.227\log_{10}q)^2\right]\dot M_{\rm disc}.
\label{eqn:jdot2}
\end{eqnarray}

Common envelope formation is expected to strongly affect the appearance of the binary, as it  rapidly becomes optically thick. By applying the Thomson cross-section to a common envelope with an exponential density profile of scale height $R_\rho$, we find that its optical thickness is:
\begin{equation}
\tau_{\rm CE, Thomson}=3.15\cdot\dfrac{(M_{\rm CE}/10^{-11}\,M_\odot)}{(R_{\rho}/10^5\,\textrm{km})^2}\dfrac{1}{\mu_{\rm e}}
\end{equation}
For example, if a WD--NS binary with a $0.15\,M_\odot$ white dwarf were to lose material into a common envelope at the Eddington rate, and assuming $R_\rho=3a$, where $a$ is the binary separation,  
the common envelope would become optically thick in slightly less than five hours.

\subsubsection{Phase 4: Energy-driven outflow}

In the case where the jet does not penetrate the envelope we consider the
energy balance.  For a common envelope of radius $R_{\rm CE}$ and efficiency
parameter $\alpha_{\rm CE}$, then, assuming a smooth flow of material out of the
common envelope, the rate at which mass is ejected by a luminosity $L$ incident
on the common envelope is
\begin{equation}
\dot M_{L} = -\frac{\alpha_{\rm CE}R_{\rm CE}L}{G(M_{\rm WD}+M_{\rm NS})}.
\label{eqn:mdotce}
\end{equation}

For our standard model we take $\alpha_{\rm CE}=1$ and $R_{\rm CE}=1.1\,a$. The choice of $\alpha_{\rm CE}$ implies that the energy used to eject the common envelope material would be equal to the decrease in the orbital energy of the binary if the ejection was due to the torques. The energy flow into the common envelope corresponds to the accretion luminosity:
\begin{equation}
L = \eta_{\rm acc}\dot M_{\rm Edd}c^2.
\label{eqn:edotce}
\end{equation}

If this rate of mass loss is greater than the flow of mass into the common
envelope then the envelope is blown off, carrying away angular momentum as
before.

\subsubsection{Phase 5: Persistent common envelope}

If the energy  input from Equation~\ref{eqn:edotce} is insufficient to drive off
all of the flow of mass in the disc wind then a persistent common
envelope builds up around the system.  We assume that this exerts a viscous drag
force on the binary and causes it to spiral inwards, ejecting the envelope.  We
parametrise the characteristic timescale for the inspiral $\tau_{\rm CE, visc}$ as
$N_{\rm CE}$ orbital periods so that the rate at which the viscous drag removes
mass from the common envelope is
\begin{equation}
\dot M_{\rm CE} = -\frac{M_{\rm CE}}{N_{\rm CE}P_{\rm orb}}.
\label{eqn:tauCE}
\end{equation}

Using the same energy balance argument as above this converts to an additional loss of
angular momentum as
\begin{equation}
\dot J_{\rm CE} = -J_{\rm orb}\frac{M_{\rm		CE}a}{\mu\alpha_{\rm CE}R_{\rm CE}N_{\rm CE}P_{\rm orb}}.
\label {eqn:JdotCE}
\end{equation}

\subsection{Typical evolution: stable and unstable systems}

\begin{figure}
\includegraphics[width=\columnwidth]{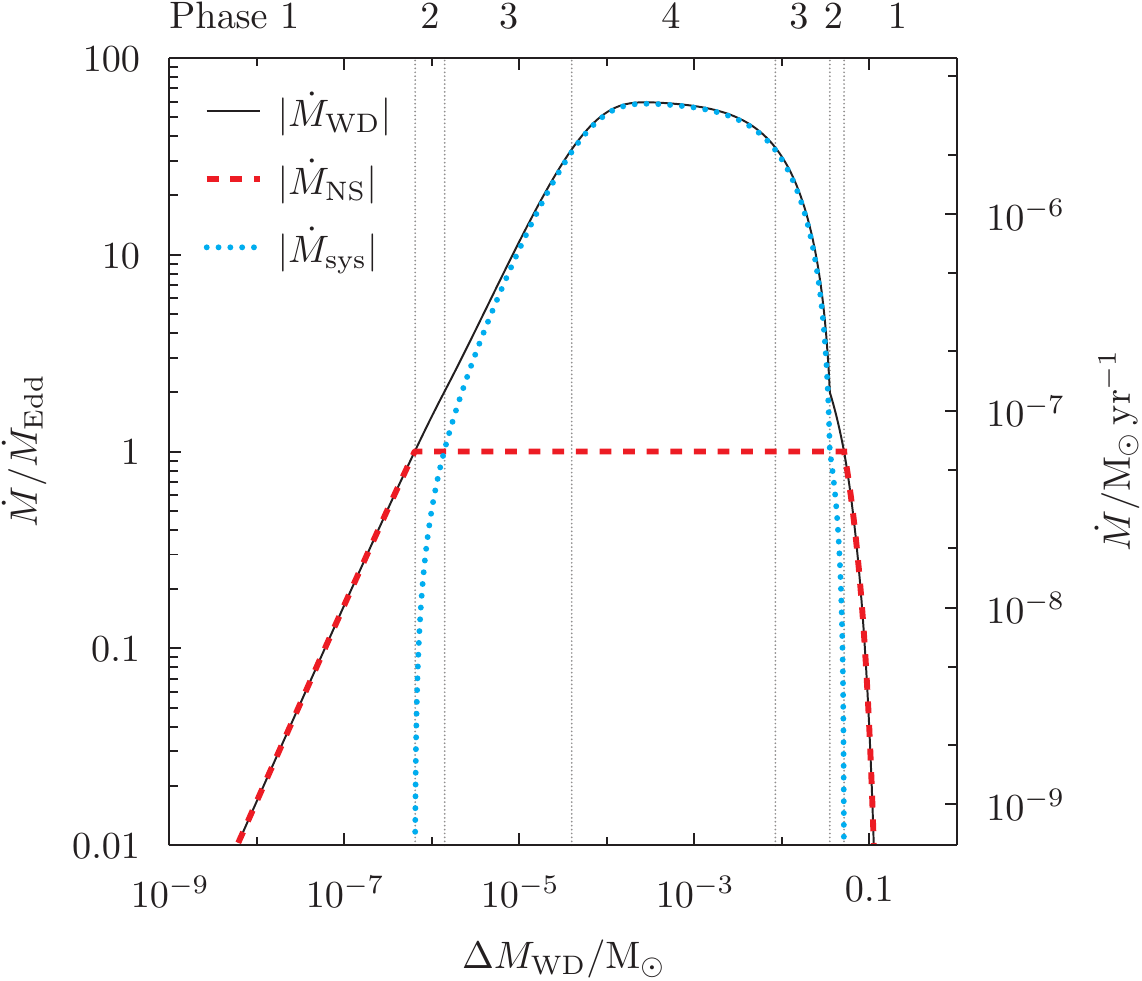}
\caption{
Evolution of a binary that initially contains a white dwarf of mass $M_{\rm
WD}=0.15\,M_\odot$ and a neutron star of mass $M_{\rm NS}=1.4\,M_\odot$. Mass-transfer rates in units of the Eddington rate $\dot M_{\rm Edd}$ (left-hand axis) and physical units (right-hand axis) are plotted as functions
of the total mass lost by the white dwarf, $\Delta M_{\rm WD}$.  The thin, solid
black line shows the rate at which the white dwarf is losing mass, the dashed
red line the rate of accretion on to the neutron star, and then dotted blue line
the rate at which mass is lost from the system as a whole.  The thin dotted
vertical grey lines separate different evolutionary phases as defined above and
labelled above the plot. The duration of each phase may be obtained from
Figure~\ref{fig:timesStandard}.}
\label{fig:mDotStable}
\end{figure}

We present first the evolution of a stable system; that is, one where the mass-transfer rate does not run away.  Initially the system contains a $1.4\,M_\odot$
neutron star and a $0.15\,M_\odot$ white dwarf. Figure~\ref{fig:mDotStable} shows the rates of change
of the white dwarf, neutron star and system masses as functions of the total
mass lost by the white dwarf, $\Delta M_{\rm WD}$.  Initially the stars are
driven together by angular momentum loss owing to gravitational radiation, but
the mass-loss rate from the white dwarf is small so the system is conservative
(phase 1).  As the mass-transfer rate increases the system goes through stages 2
and 3 into stage 4 in about $10$ years time. The mass transfer becomes highly non-conservative, with a
rate of over $200$ times the neutron star's Eddington accretion rate, but at all
times the energy input into the common envelope is sufficient to eject the
mass that is being transferred. Stage 4 lasts for about $10^{4}$ years. Ultimately, despite the loss of angular momentum in the
mass loss from the accretion disc, the orbit widens and the mass-transfer rate
goes down again.  At the point, when the system re-enters conservative evolution after several hundred thousand years of non-conservative evolution  in phases 2, 3 and 4, the white dwarf has a mass of just under $0.1\,M_\odot$. The subsequent
evolution is of a low-mass white dwarf transferring mass to a neutron star, and is compatible with ultra-compact {\it X}-ray binaries such as 4U\ 1820-30.  

\begin{figure}
\includegraphics[width=\columnwidth]{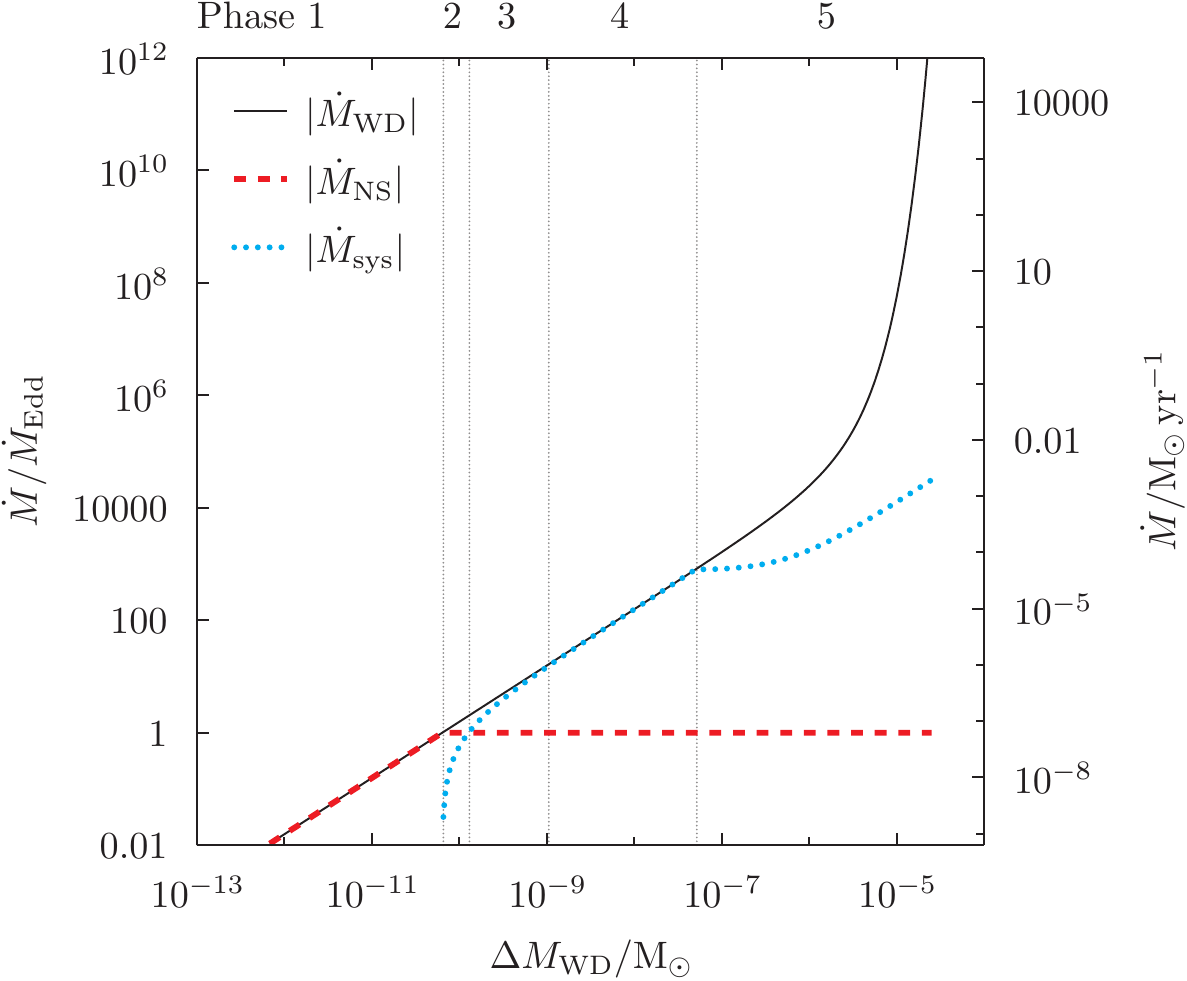}
\caption{As in Figure~\ref{fig:mDotStable} but with a white dwarf of initial
mass $M_{\rm WD}=0.75\,M_\odot$.
}
\label{fig:mDotUnstable}
\end{figure}

In contrast, an unstable system exhibits an exponentially increasing rate of
mass transfer as predicted by the classical stability analysis.
Figure~\ref{fig:mDotUnstable} shows the same quantities as 
Figure~\ref{fig:mDotStable} but for a system where the initial mass of the white
dwarf is $0.75\,M_\odot$.  The rate of mass loss from the system does not turn
over but instead increases, and the non-conservative phase lasts for only about one day.  We terminate the evolution once the timescale for
the evolution of the orbital angular momentum reduces to within ten orbital
periods. At this point the white dwarf will be tidally shredded by the neutron star, and the evolution will take place on dynamical timescales.  

\subsection{Time spent in different phases}

\begin{figure}
\includegraphics[width=0.85\columnwidth]{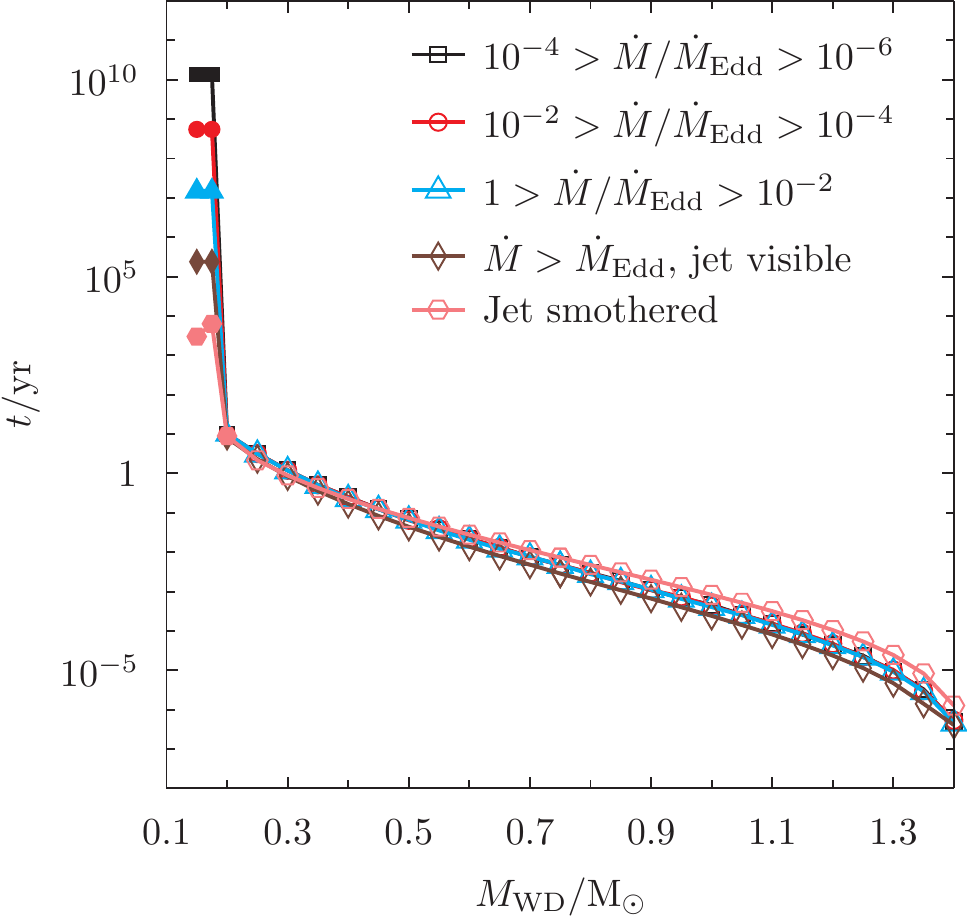}
\caption
{
Time for which the binary is in different evolutionary stages, using our
standard assumptions.  The time $t$ in years is plotted as a function of the
initial white dwarf mass $M_{\rm WD}$ in solar masses.  Systems that evolved to
long-lived stable configurations are plotted with solid symbols; systems that
merge with open symbols.  Black lines with squares show the time spent accreting
conservatively 
between $10^{-6}$ and $10^{-4}$ of the Eddington rate $\dot M_{\rm Edd}$, red lines with circles
the time spent between $10^{-4}$ and $10^{-2}$ of $\dot M_{\rm Edd}$, and blue
lines with triangles the time spent between $10^{-2}\,\dot M_{\rm Edd}$ and $\dot M_{\rm Edd}$.  The brown line with diamonds shows the time spent accreting at super-Eddington rates with the jet visible, and the pink line with hexagons the time when the jet is smothered by the common envelope.
}
\label{fig:timesStandard}
\end{figure}

For the purposes of considering the effects of our results on potential
observations, we plot the time spent in various states as a function of
initial white dwarf mass.  The results are shown in
Figure~\ref{fig:timesStandard}.  The clearest distinction is between stable
systems, with $M_{\rm WD}<0.2\,M_\odot$, and the unstable systems.  The former
have much larger times in all stages of evolution.  This is because the evolution after the peak in mass-transfer rate is much slower than the evolution before the peak. The duration of the lowest accretion rate phase
($10^{-4}>\dot M/\dot M_{\rm Edd}>10^{-6}\,$, black squares) is limited by us
stopping the simulations once a Hubble time has expired.  In contrast, for the
unstable systems by definition only the evolution before the peak mass-transfer
rate can be observed.  Each of the phases has roughly the same duration for a
given initial white dwarf mass, which comes about because in the conservative
and jet phases the orbital semi-major
axis $a$ and the degree of overflow $\Delta R$ are changing at a roughly constant rate with time, and so the mass-transfer rate changes exponentially with time following Equation~\ref{MDotInst}.
It can be seen, moreover, that these times are all rather short.  Even for the
white dwarfs which are just above the unstable-stable boundary, the time between the onset of significant mass transfer and final merger is only a few years.

\subsection{The effect of our assumptions on the stability boundary}
\label{section:stab}

If one is interested either in the formation rates of UCXBs -- i.e. the stable
systems -- or of transient events formed from the merger of unstable systems, the
crucial question is the location of the stability boundary. The most important factor influencing stability is the amount of angular momentum carried off by the material lost into the disc \citep{Webbink1985}. A greater loss of angular momentum
shrinks the orbit more quickly and hence pushes the binary towards instability.

\begin{figure}
\includegraphics[width=0.85\columnwidth]{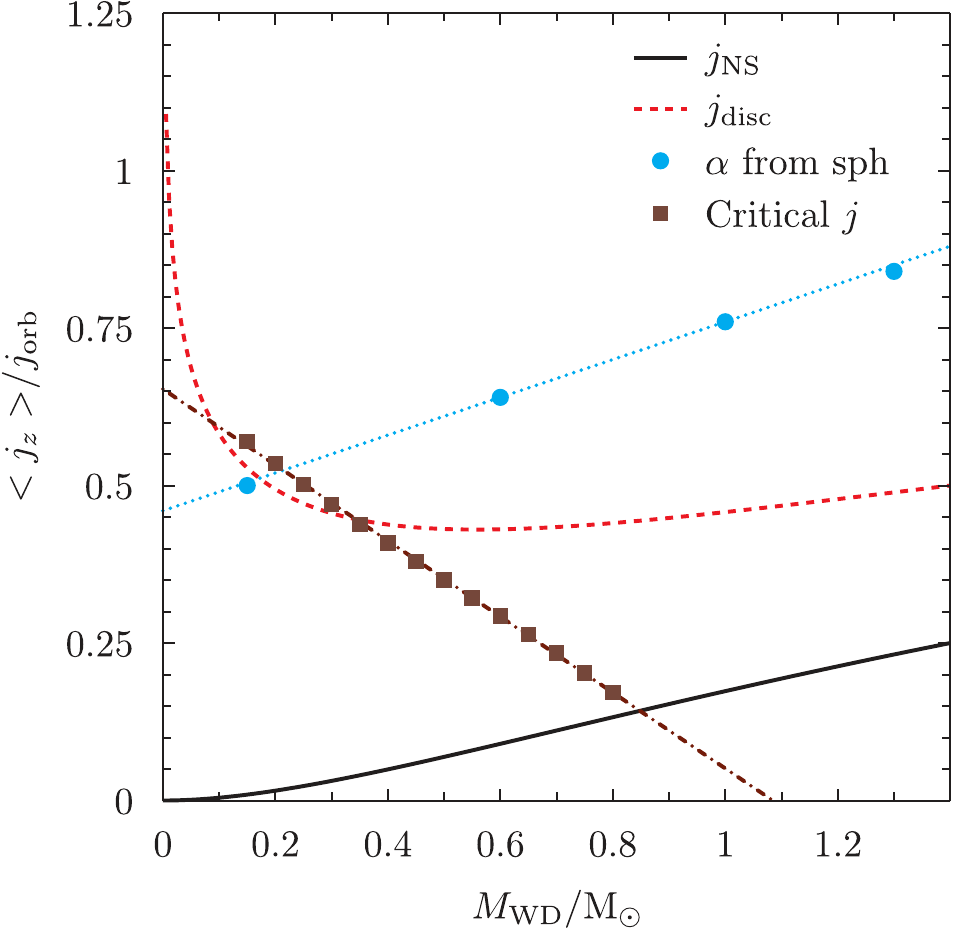}
\caption
{
Different estimates of the specific angular momentum lost from the system, $<j_z>$,
in units of the specific orbital angular momentum $j_{\rm orb}$, as a function
of white dwarf mass $M_{\rm WD}$.  The black solid line shows the specific
orbital angular momentum of the accreting neutron star, $j_{\rm NS}$.  The red
dashed line shows the specific orbital angular momentum of a disc orbiting the
accreting neutron star at the circularisation radius.  The blue filled circles
show the specific orbital angular momentum lost in the mechanical wind in our
SPH simulations, and the dotted line a straight-line fit. The SPH estimate is subject to some uncertainty and may be up to $20$ per cent higher, depending on the criteria by which the lost material is associated with the common envelope. The brown squares and best-fit dot-dashed line show the critical specific angular momentum that
separates stable and unstable systems.
}
\label{fig:angMomTest}
\end{figure}

Figure~\ref{fig:angMomTest} shows three different prescriptions for the angular momentum loss, quantified by the $\alpha$, which measures the specific angular momentum carried away by the lost material. The brown line shows the critical specific angular momentum for each initial mass. A greater loss of angular momentum corresponds to unstable mass transfer. The critical line is obtained by keeping the specific angular momentum of the material lost in the mechanical disc wind constant and finding the value which lies on the borderline of stability. The blue dotted line corresponds to the model in which disc winds carry away the angular momentum, as measured in our SPH simulations and used in our standard conditions for the long-term evolution (Equations~\ref{AlphaFit}~and~\ref{eqn:jdot1}). The red dashed line in the figure corresponds to the model in which disc winds carry away the material with the specific angular momentum at the circularisation radius around the neutron star (Equation~\ref{eqn:jdot2}). The black line represents the mass loss model where the material is lost entirely through a jet and carries away the specific angular momentum of the neutron star.

As one may see from Figure~\ref{fig:angMomTest}, for low-mass donors the disc winds are much more efficient than jets at removing angular momentum from the orbit. This is expected, since material lost in the jet carries off the specific angular momentum of the neutron star.  For low-mass donors the binary centre of mass is close to the neutron star and hence it has a very small orbital angular momentum.   The disc, on the other hand, may retain significant angular momentum.  The value depends on the viscous timescale for the disc, which we find to be relatively short. Hence we expect that material will be ejected from the disc before disc instabilities and the resulting torques can return its angular momentum to the binary. Similarly, the neutron star cannot be an efficient sink of angular momentum due to its small physical size. Therefore, we expect the specific angular momentum of material lost in the wind to be close to typical values in the disc. Because material lost in a jet carries off relatively little angular momentum, the jet-only mass-loss model corresponds to slower orbital evolution at a given mass and hence a much higher critical mass than we measure when including the effects of disc winds.

In our SPH simulations the material carries away more angular momentum than that of a disc at the circularisation radius at all but the lowest masses.
At lower white-dwarf masses, the white dwarf has less dynamical effect on the wind as it leaves the disc.  Reassuringly, in the SPH runs with the lowest-mass white dwarf we measure that the angular momentum carried off is roughly equal to that of the disc at the circularisation radius.  At higher white dwarf masses, however, the binary motion increasingly influences the outflowing material, imposing an additional torque as it escapes the disc. The lost material therefore gains additional angular momentum.

Our SPH estimates of $\alpha$ are somewhat sensitive to the choice of the boundary of the common envelope, shown in Figure~\ref{fig:MTScheme}. We test this by varying the radii of the surfaces that divide the particles associated with the stars from those associated with the common envelope, $R_{\rm div}$, from $1.4\,R_{\rm RL}$ to $1.8\,R_{\rm RL}$, which leads to an increase in $\alpha$ by about $20$ per cent  for all of our models. For low-mass helium white dwarfs we expect the SPH values of $\alpha$ to be close to the ones corresponding to the circularisation radius, as argued above. Yet, even if the correct values of $\alpha$ were higher, we can still form observed UCXB systems such as 4U\ 1820-30. 
An increase in $\alpha$ for the binaries with higher-mass CO or ONe white dwarf donors will have no significant effect on their evolution since these systems are already unstable.  Regardless of the exact value of $\alpha$, for the range of plausible values the {\it qualitative} behaviour is the same.  
Stable mass transfer is possible only for the lowest-mass He white dwarf donors.  Other systems, including all C/O and O/Ne white dwarf systems, will undergo unstable evolution and merge.

The other model parameters have relatively little influence on the location of the stability boundary, as can be seen in Table~\ref{tab:sensitivityTests}.  The viscous drag on the orbit owing to the presence of a common envelope makes the
system less stable as the viscous inspiral extracts further angular momentum
from the orbit. Similarly the jet removes mass with smaller specific angular momentum and hence larger values of $f_{\rm jet}$ make the system more stable. Overall, the main parameter determining the stability boundary is $\alpha$, i.e. the measure of the angular momentum carried away by the lost material. Our model predicts that the stability boundary is between $0.2$ and $0.3\,M_\odot$, largely independently of the choice of the model parameters.

\begin{table}
\caption{Results of different binary models}
\begin{tabular}{p{.7\columnwidth}l}
\hline
Settings & Critical mass [$M_\odot$] \\
\hline
Defaults: $f_{\rm jet}=1$, $j_{\rm wind}=j_{\rm fit}$, $R_{\rm CE}=1.1\,a$, $N_{\rm CE}=10^4$ & 0.199 \\
$j_{\rm wind}=j_{\rm circ}$      & 0.262 \\
$j_{\rm wind}=j_{\rm jet}$       & 0.844 \\
\hline
$N_{\rm CE}\rightarrow\infty$    & 0.202 \\
$N_{\rm CE}=100$                 & 0.192 \\
$R_{\rm CE} = 3\,a$              & 0.200 \\
\hline
$f_{\rm jet} = 0.01$             & 0.199 \\
$f_{\rm jet} = 10$               & 0.200 \\
\hline
\end{tabular}
\label{tab:sensitivityTests}
\end{table}

%When using the alphas based on circularisation radius estimate as defaults:
%Defaults: $f_{\rm jet}=1$, $j_{\rm wind}=j_{\rm disc}$, $R_{\rm CE}=1.1\,a$, $N_{\rm CE}=10^4$ & 0.262 \\
%$j_{\rm wind}=j_{\rm fit}$       & 0.200 \\
%$j_{\rm wind}=j_{\rm jet}$       & 0.844 \\
%\hline
%$N_{\rm CE}\rightarrow\infty$    & 0.329 \\
%$N_{\rm CE}=100$                 & 0.262 \\
%$R_{\rm CE} = 3\,a$              & 0.286 \\
%\hline
%$f_{\rm jet} = 0.01$             & 0.262 \\
%$f_{\rm jet} = 10$               & 0.264 \\

\subsection{Observational implications of our modelling}

\label{sec:DiscObs}

Our models imply that the critical white-dwarf  mass that divides stable from unstable systems is significantly lower than predicted by the commonly assumed jet-only model of mass loss.
This constrains the formation channels of UCXBs, since they only form  following stable mass transfer. Our model implies that UCXBs with observed excesses of C, O or Ne can only have formed from close binaries containing a helium star and a neutron star.  This is significant for the spectroscopic donor diagnostic developed in \citet{Nelemans2010}. Helium-transferring systems, which 4U~1820-30 is thought to possibly be, may still form from the binaries with low-mass helium white dwarf donors. 

Applying our analysis to the binaries with He star donors and assuming the mass-radius relation $R/R_{\odot}=0.2(M/M_{\odot})^{0.86}$ from \citet{DeLoore1992}, we find that for donor masses below $1.2\,M_\odot$ the mass transfer remains sub-Eddington and hence stable. Unlike for WD donors, the critical mass for helium star donors is sensitive to the mass-transfer rate at which the disc wind is launched. The donors with mass of $1.65\,M_\odot$ will reach a maximal rate of $2\,\dot{M}_{\rm Edd}$ when on stage 2, while the donors with mass of $2.15\,M_\odot$ will reach a rate of $10\,\dot{M}_{\rm Edd}$. Since the disc wind efficiently destabilizes the system the critical mass is expected to belong to this interval of masses.

Our results may have some relevance for population synthesis of UCXBs. For example \citet{vanHaaften2013} predict in their standard model $35$~--~$50$ bright UCXBs in the galactic bulge, whereas only $3$ are supposedly observed there \citep{2012A&A...543A.121V}. At the same time, they report having $97.4$ per cent of WD donors in UCXBs having masses above $0.38\,M_\odot$. Removing the unstable systems from their population makes it better match the observations.

Once old UCXBs reach a sufficiently low mass-transfer rate, their accretion discs will become tenuous, radiatively-inefficient and geometrically thick \citep{Narayan2008}. If this is accompanied by disc-wind mass loss, the systems enter the unstable regime as follows from the low-mass end of Figure~\ref{fig:angMomTest}. If this is indeed the case, it may provide another explanation for the lack of observed UCXBs with donor masses below $0.01\,M_\odot$, discussed e.g. in \citet{Lasota2008}, where the effects of irradiation on the disc were studied.

We now use our values for the critical mass to associate outcomes with the populations of WD--NS binaries discussed in Section~\ref{PopObs} and summarised in Figure~\ref{fig:PopulationsChannels}. Field population one, where the neutron star is formed after the white dwarf, for example, is expected to be the dominant source of WD--NS derived transients in the field. Field population two, where the neutron star is formed before the white dwarf, is expected to be the main progenitor for UCXBs in the field. In globular clusters most UCXBs are expected to come from the binaries formed through dynamical collisions of neutron stars and red giants (population three) whereas most transient events are expected to come from the binaries formed through dynamic exchanges (population four).

We test that the empirically observed formation and merger rates of WD--NS binaries and products of their evolution are broadly consistent with our conclusions, as summarised in Table~\ref{table:EmpiricalMergerRates}. Empirical merger rates for detached binaries are based on binary pulsar systems which have a white dwarf companion and will merge in less than a Hubble time as provided by the ATNF pulsar catalogue. We used the pulsar characteristic lifetimes, which were rounded upwards to nearest power of ten megayears (for example, $\tau_{\rm ch}=300\,\rm{Myr}$ would be rounded to $1000\,\rm{Myr}$). The rounding decreases the inferred merger rates of binary pulsars and was done in order to reduce the sensitivity to the uncertainties in the ages of young pulsars such as PSR~J1141-6545 \citep{Kim2004}. As such, cautious lower bounds are set on the highly uncertain merger rates. The observational bias for field systems $f_{\rm b}=400$ was adopted from \citet{Kim2003}, and the beaming bias $f_{\rm beam}=6$ is from \cite{Kalogera2001}. The empirical merger rates for UCXBs is based on the list of confirmed UCXBs from \citet{2012A&A...543A.121V}. Their typical visibility lifetime of $100\,\textrm{Myr}$  corresponds to them having luminosities above $10^{36}\,\textrm{erg}/\textrm{s}$ and thus being visible in all-sky surveys (RXTE/ASM,
SWIFT/BAT). The empirical occurrence rate for Ca-rich gap transients is based on the list of nine events in \citet{Lyman2014} which have a known host with existing distance measurement. Converting from the typical observed event rate of  $1\,\textrm{yr}^{-1}$ to the galactic rate is done by assuming extragalactic blue luminosity density of $(1.98\pm0.16)\times 10^{8} L_{\rm \odot,B}/\textrm{Mpc}^3$, as from \citet{Kopparapu2008}, and the Milky way blue luminosity of $1.7\times 10^{10} L_{\rm \odot,B}$, as from \citet{Kalogera2001}. The typical value for the distance horizon was set to $50\,\textrm{Mpc}$, the typical distance to the observed gap transients.

As one may see from Table~\ref{table:EmpiricalMergerRates}, the merger rates from the WD--NS systems containing ONe white dwarf appears to match those for Ca-rich gap transients. The empirical merger rate for these systems exceeds significantly the corresponding rates for the systems with a He or CO white dwarf. The main source of the difference are short characteristic lifetimes of less than $100\,\textrm{Myr}$ of two field binary pulsars PSR J1141-6545 and PSR J1952+2630. It is also very unlikely observationally that any interesting fraction of ONe systems can contribute to UCXB formation. The fact that the empirical merger rate for the binaries with a He WD exceeds the UCXB formation rate may possibly be explained by a fraction of WD--NS binaries undergoing ablation from the pulsar and never coming into to contact. Another possible mechanism is ablation due to ongoing mass transfer, which may shorten the lifetime of UCXB systems \citep{vanHaaften2012}.

\begin{table}
\caption{Empirical merger rates for observed systems}
{\setlength\tabcolsep{1.5pt}
\begin{tabular}{ccccc}
\hline
System type & NS-(He WD) & NS-(CO WD) & NS-(ONe WD) \\ 
\hline
Emp. rate, $\textrm{Myr}^{-1}$ & $0.72$ & $0.24$ & $260$ \\
Objects & $3$ & $1$ & $2$ \\ 
\hline
System type & He-only UCXBs & All UCXBs & Ca-rich gap trans.\\
\hline
Emp. rate, $\textrm{Myr}^{-1}$ & $0.05$ & $0.13$ & $160$ \\ 
Objects & $5$ & $13$ & $9$ \\ 
\hline
\end{tabular}}
Notes: Empirical merger rates (per MW-like galaxy) as inferred from the population of binary radio pulsars (top half), or the UCXBs and Ca-rich gap transients (bottom half). Our model predicts that empirical formation rates for He-only UCXBs should be close to empirical merger rates for NS-(He WD) binaries. CO or ONe UCXBs, instead, are predicted to form entirely from a separate channel of (NS-He star) binaries. Similarly, if WD-NS systems are progenitors for Ca-rich gap transients, we expect the empirical merger rates for NS-(ONe WD) binaries to match the empirical occurrence rates of Ca-rich gap transients. Both these predictions appear consistent with the empirical merger rates we obtain. We list the numbers of objects used to make the inference. The data for binary pulsars is obtained from ATNF pulsar catalogue. NS-He WD binaries include the systems with the median companion mass above $0.08\,M_\odot$. NS-(He WD), NS-(CO WD) and NS-(ONe WD) binaries are identified with the observed binary pulsars based on the median donor mass, unless the mass is known from spectroscopic observations. He-only UCXB systems are identified among all the UCXB systems by the absence of C or Ne signatures in their spectra.
\label{table:EmpiricalMergerRates}
\end{table}

We further estimate if any unstable systems may be visible to the all-sky {\it X}-ray survey eROSITA before they merge. Using the expected sensitivity for point sources of $1.1\times 10^{-14}\,\textrm{erg}/(\textrm{cm}^2\cdot\textrm{s})$, as provided in \citet{Merloni2012}, the horizon distance for seeing the sources at the Eddington luminosity is $16.4\,\textrm{Mpc}$. Using the typical merger rate of $100\,\textrm{Myr}^{-1}$ for unstable systems and assuming that {\it X}-ray radiation is beamed in an angle of $11.5^\circ$, one expects $0.8$ sources to be visible on the sky over the four-year duration of the mission. However, these sources are not expected to be observed in the field of view of the instrument, since they evolve through the bright phase on short timescales of several hours, as may be seen from Figure~\ref{fig:timesStandard}. By using the merger rates for helium UCXBs or for binary pulsars with He white dwarf, one may conclude that eROSITA will observe, correspondingly, between $50$ and $750$ Eddington luminosity sources evolving through a stable phase. It will be impossible, however, to distinguish these systems from LMXB and HMXB sources having other evolutionary pathways which happen to be in the same {\it X}-ray luminosity band.

\section{Conclusions}
\label{sec:ConclusionsMain}
 
We have developed a modified smoothed particle hydrodynamics code and applied it to model mass transfer in circular and eccentric binaries containing a white dwarf and a neutron star. Using these simulations, we have determined the efficiency with which disc winds remove angular momentum from the binary when it is undergoing highly super-Eddington mass transfer. Based on these results we have constructed a model for the long-term evolution of WD--NS binaries after they start to transfer mass.  We use the model to determine which systems undergo stable mass transfer and evolve into ultra-compact {\it X}-ray binaries, and which undergo unstable mass transfer and potentially give rise to luminous transients. 

Our work builds on studies of mass transfer in eccentric binaries at realistic mass-transfer rates by \citet{Church2009} and \citet{Lajoie2011a}. We have incorporated recent developments in SPH summarized in \citet{Rosswog2014}, and also improved the treatment of mass transfer, as summarised in Section~\ref{SPHOWat}. We show that the instantaneous mass transfer rate depends on how many scale heights the Roche lobe digs into the atmosphere of the donor star at pericentre. This simple treatment allows us to reproduce the mass-transfer rates observed in simulations, and hence derive the appropriate mass-transfer rate for all realistic WD--NS binaries.  Our results may inform the detailed analytic studies of secular evolution of mass transfer in eccentric systems \citep{Dosopoulou2016a, Dosopoulou2016,Sepinsky2009, Sepinsky2007} as we measure orbital profiles of mass transfer on eccentric orbits. In agreement with \citet{Lajoie2011}, for example, we find that the peak of mass-transfer rate lags behind the periastron passage for non-vanishing eccentricities.

A fraction of WD--NS binaries will come into contact with sufficiently large eccentricities that mass transfer will turn on and off during each orbit. Our simulations suggest that the eccentricity most likely does not affect the long-term evolution of these binaries. The rate of loss of angular momentum from the accretion disc around the neutron star, which determines the final outcome of the mass transfer, is not affected by eccentricity because the transferred material spends sufficient time in the disc before being lost as the viscous timescale is somewhat
longer than the orbital period. The eccentricity does change the mass-transfer rate at a given orbital separation, but our work implies that the effect of the eccentricity may be factored out from the mass loss law. This implies that the long-term evolution will only be affected by eccentricity if it changes on very short timescales comparable to those of mass-transfer rates.

The standard model of angular momentum loss, widely used in the literature, assumes that the accreted material is lost in a jet from the neutron star and hence carries off its specific orbital angular momentum.  At highly super-Eddington mass-transfer rates, however, material cannot be efficiently ejected from the immediate vicinity of the accreting neutron star, so disc winds are expected become important. These mass-transfer rates are expected to be reached by almost all the WD--NS binaries \citep{2012A&A...537A.104V}, and so a correct treatment of this phase is necessary to determine the critical mass that separates stable and unstable evolution.  We find that for the lowest-mass donor white dwarfs the disc winds carry off the specific angular momentum of a disc formed at the circularisation radius of the transferred material.  At larger white dwarf masses, where the orbital motion has a larger dynamical effect, we measure in
our simulations  a wind that carries away a somewhat greater specific angular momentum.

Based on our measurements of specific angular momentum in the disc winds, we find that the critical donor mass for mass transfer to be unstable is about $0.2\,M_{\odot}$.  This is much lower than the critical mass of
 about $0.84\,M_{\odot}$ obtained from the jet-only model. The difference is explained by the fact that disc winds are much more efficient at removing angular momentum from the binary than jets.   
While the accretor has a low specific angular momentum, especially for low-mass donors, the angular momentum in the disc orbiting the accretor is still significant. Once the disc winds are launched, the systems lose more angular momentum which tightens the orbit and hence increases the mass-transfer rate.  Thus the higher rates of angular-momentum loss lead to a reduction in  the stability of an individual system, thus reducing the critical mass.

We test whether our conclusions are consistent with observations of WD--NS binaries  and their progeny.  The critical mass that we derive implies that only He white dwarfs can undergo stable mass transfer and form UCXBs. This implies that C and O rich UCXBs must be descended from helium-star donors (rather than
white dwarfs).  All CO and ONe white dwarfs are expected to lead to transient sources, potentially Ca-rich gap transients. We infer the expected galactic merger rate from the lifetimes and merger rates of observed binary pulsars with CO and ONe white dwarfs.  The derived rate is consistent with the apparent rate of of Ca-rich gap transients, and is rate dominated by systems with ONe white dwarfs. We have investigated whether the precursor systems should be visible in {\it X}-rays in the earlier stages of their inspiral, however, this appears unlikely with currently-planned instruments.

Our updated Oil-on-Water technique may be applicable to other studies modelling mass transfer on timescales that are long compared to the binary's orbital period.  We show that the effects of quadrupole interactions and donor radius expansion need to be taken into account in setting up the the initial orbit for these binaries; we expect that this will be true for most hydrodynamic simulations of interacting binaries.  We have verified numerically that the widely-used formula of \citet{Ritter1988} applies to WD--NS binaries.  The approximation of instantaneous donor adjustment may possibly hold for other types of eccentric systems with different donors. Finally, the overall method of incorporating results from hydrodynamic simulations carried out on orbital timescales with long-term analytic calculations may have a wider application to the evolution of other classes of binary systems.

\section*{Acknowledgements}

The authors would like to thank Christopher Tout, Lennart Lindegren, Lennart van Haaften, Ralph Wijers, Brian Metzger, Francis Timmes and the anonymous referee for helpful discussions and comments. R.P.C. was supported by the Swedish
Research Council (grants 2012-2254 and 2012-5807). M.B.D. was  supported  by  the  Swedish  Research  Council (grants 2008-4089 and 2011-3991). The simulations were performed on the resources provided by the Swedish National Infrastructure for Computing (SNIC) at the Lunarc cluster.

\appendix

\section{SPH Method Developments}
\label{SPHMain}
Throughout this Section we follow the notation from Appendix~A1 of \citet{2010MNRAS.408..669C}.
\subsection{$\rho-h$ relation}
\label{SPHRhoH}

We use a $\rho-h$ scheme which defines a maximal size for SPH particles, which we set equal to $R_{\rm WD}$. For particles much smaller than the limiting radius the formulation is identical to the standard one. For larger particles, the radius is limited in a Lagrangian conservative way consistent with the SPH equations, which helps to treat the fall-back problem described in Section~\ref{SPHOWat}. Comparing to the standard approach, where the maximal size of SPH particles or the maximal number of neighbours is enforced, our method provides correct accelerations and heating terms for particles of all sizes whilst being computationally efficient. The required modifications to our SPH code are minimal.

The defining equations, $h_i^\nu\hat{\rho}_i=\sum_j m_j w(r_{ij})$ and $h_i^\nu(\hat{\rho}_i+\hat{\rho}_0)=M_h$, follow the idea from \citet{Monaghan2005}, their equation (4.5). Here $h_i$ is the smoothing length for particle $i$, $\hat{\rho}_i$ is its density estimate, $m_i$ is its mass, $\nu$ is the number of dimensions, $w(r)$ is the kernel function, $r_{ij}$ is the distance to particle $j$ and $M_h$ is a variable used to control the average number of neighbours. $\hat{\rho}_0$ defines the maximal size of SPH particles, which we set equal to the radius of the donor star, and the standard formulation is recovered for $\hat{\rho}_0\ll\hat{\rho}_i$, which holds in all but the outermost low-density regions. 

The only SPH equations which require a modification are those in the $\rho-h$ iteration loop:
\begin{equation}
h_i\leftarrow h_i\left(\dfrac{M_h}{h_i^\nu(\hat{\rho}_i+\hat{\rho}_0)}\right)^{f_i/\nu},\quad f_i=-\nu\dfrac{\sum_j m_j w_{ij}+h_i^\nu\rho_0}{\sum_j m_j r_{ij}^2 \tilde{w}_{ij}-\nu h_i^\nu\rho_0},
\label{SPHRhoHIter}
\end{equation}
where $\tilde{w}(r)\equiv w^\prime(r)/r$.
The convergence properties are better than in the standard case since the presence of the $\rho_0$ term prevents an arbitrary growth of $h_i$ in the low density regions.  

This modification improves the treatment of the particles falling back onto the white dwarf, preventing high computational costs or non-physical jumps in the energy, density or scale height of these particles. We note that this modification does not affect the dynamics of the disc or the disc winds, nor any of the properties measured and presented in Section~\ref{sec:ResMain}.

\subsection{Oil-Water forces}
\label{SPHOWForce}

Below we summarize the updated numerical treatment  for Oil and Water interaction. It is implemented with the aim of defining a smooth, impenetrable boundary between Oil and Water particles which is based on conservative forces and is consistent with a Lagrangian SPH formulation.

Water particles experience gravity, pressure gradient  forces and viscosity from other Water particles. Oil particles experience pressure gradient forces and viscosity from Oil particles. Additionally, they experience gravity forces from Water particles and Oil-Water forces offsetting the gravity.

The Oil-Water forces are computed according to:
\begin{equation}
({\bf\dot{v}}_i)_{{\rm OW}}= -K\sum_{j,wat}\dfrac{P_j}{\hat{\rho}_j^2}\dfrac{f_j}{h_j^{\nu+1}}\tilde{w}^{{\rm OW}}_{ji}{\bf x}_{ji},
\end{equation}
where $\tilde{w}^{{\rm OW}}(r)\equiv w^{{\rm OW}\prime}(r)/r$, $P_j$ is the pressure of the particle $j$ and $K$ is set to $16$. Gravity forces acting on Oil particles from Water particles follow standard SPH equations.

The force expression is calculated from an interaction Lagrangian. The equipotential surfaces for it follow the surfaces of constant pressure for Water, and the Oil-Water kernel $w^{{\rm OW}}$ defines the field strength. The method ensures that the Oil-Water force field is smooth and conservative, which allows for higher particle density contrasts.

The kernel $w^{{\rm OW}}(r_{ji})$ is set equal to the SPH kernel $w_{ji}$ for $r_{ji}>0.3$. For $r_{ji}<0.3$ the expression is matched $C_2$ continuously to $A+Br_{ji}+C/r_{ji}$ by choosing appropriate $A$, $B$, $C$. The idea of introducing an infinite potential wall defined by $w^{{\rm OW}}$ is inspired by \citet{2009CoPhC.180.1811M}. The treatment ensures that Oil particles do not penetrate through the Water surface.

\section{Physical motivation for the instantaneous response model}
\label{MDotMotivation}

In our simulations, the mass-transfer rates on eccentric orbits  follow the expressions for circular rates calculated instantaneously to a good approximation. We show here that this is expected physically, since the material is transferred from a small surface region of the donor, and derive criteria for the approximation to work. 

During mass transfer the material outside the Roche lobe of the donor is gravitationally unbound. In standard mass-transfer models, e.g. \citet{Ge2010}, the material flows towards $L_1$ along gravitational equipotential surfaces. In reality the unbound material has to be replenished. Most of it crosses the Roche surface through a region centred around $L_1$ and having a characteristic size $L_{\rm MT}$.

The mass inflow rate is equal to $\dot{M}_\perp = \rho_{\rm RL,WD} (\pi L_{\rm MT}^2)\langle v_\perp \rangle$.
The density $\rho_{\rm RL,WD}$ may be determined from a $1$~D stellar model. The mean inflow speed may be parametrised through the sound speed as $\langle v_\perp\rangle = \lambda\cdot c_{\rm s,RL,WD}$, with $\lambda$  expected to be small and between $0.1$ and $0.01$. By equating this to $\dot{M}$ from the model provided in \citet{Ge2010} and assuming isothermal atmospheres, one obtains:
\begin{equation}
\dfrac{L_{\rm MT}}{R_{\rm WD}}= \dfrac{3}{2}\sqrt{\dfrac{h_\rho }{R_{\rm WD}}}\sqrt{\dfrac{1}{\lambda}}
\end{equation}
which holds for $q=M_{\rm WD}/M_{\rm NS}$ between $0.01$ and $1$, within $10$ per cent accuracy. Since $h_\rho$ is several orders of magnitude smaller than $\lambda R_{\rm WD}$, $L_{MT}$ is a relatively small fraction of the WD radius.

We further compare the relevant physical timescales to the duration of mass-transfer episode $\tau_{\rm MT}=(P/\pi)\sqrt{h_\rho/(R_{\rm WD}e)}$.
The typical time for gravity waves to travel across the transferring region is $L_{\rm MT}/v_{\rm WD,Kep}$. It is shorter than $\tau_{\rm MT}$ if $e<25\lambda$, which is likely to always hold true.
The sound crossing time is given by $L_{\rm MT}/c_{\rm s,RL,WD}$. It is smaller than $\tau_{\rm MT}$ if $e<50 \lambda (h_{\rho}/R_{\rm WD})$. The former condition is determined by the dynamical free-fall time and indicates that the model of instantaneous response is likely consistent with the measurements presented in Table~\ref{table:EccentricMeasurements}. The latter condition is necessary for a quasi-stationary flow to exist.

%\clearpage

\bibliographystyle{mnras}
\bibliography{BibList}

\bsp	% typesetting comment
\label{lastpage}

\end{document}